\documentclass[twocolumn]{aastex7}
\usepackage{graphicx}
\usepackage{xcolor}
\usepackage{graphics}
\graphicspath{{}}
\usepackage{amsmath}
\usepackage{amssymb}
\usepackage{url}
\usepackage{hyperref}

\shorttitle{CGM Emission VSFs}
\shortauthors{Bouchereau et al.}

\begin{document}

\title{Measuring Simulated Circumgalactic Medium Turbulence with Emission-Weighted Projected Velocity Structure Functions in FOGGIE}

\author{Helena Bouchereau}
\affiliation{Department of Physics, Northeastern University, Boston, MA, 02115, USA}
\email{enter email here}

\author[0000-0003-1785-8022]{Cassandra Lochhaas}
\altaffiliation{NASA Hubble Fellow; corresponding author}
\affiliation{Center for Astrophysics $|$ Harvard \& Smithsonian, 60 Garden St., Cambridge, MA 02138, USA}
\email{clochhaas@cfa.harvard.edu}

\author[0000-0002-8573-2993]{Jake S. Bennett}
\affiliation{Center for Astrophysics $|$ Harvard \& Smithsonian, 60 Garden St., Cambridge, MA 02138, USA}
\affiliation{School of Physics \& Astronomy, University of Nottingham, University Park, Nottingham NG7 2RD, UK}
\email{Jake.Bennett@nottingham.ac.uk}

\author[0000-0002-8739-3163]{Mandy C. Chen}
\affiliation{Cahill Center for Astronomy and Astrophysics, California Institute of Technology, Pasadena, CA 91125, USA}
\affiliation{The Observatories of the Carnegie Institution for Science, 813 Santa Barbara Street, Pasadena, CA 91101, USA}
\email{mandyc@caltech.edu}

\begin{abstract}
The spatially-resolved kinematics of line emission from the circumgalactic medium (CGM) of a galaxy can contain information about the CGM turbulence, which may play an important role in galaxy evolution. Due to the region’s diffuse nature, there have been limited observations of low-redshift CGM emission until recent efforts that use spatially-resolved emission line kinematics to probe CGM turbulence. We use velocity structure functions (VSFs) as a measure for the properties of turbulence using the high-resolution cosmological zoom-in FOGGIE simulations. We focus on the location of the ``turnover" in the VSF slope, often used as a measurement of the turbulence driving scale, and study how resolution, measurement area size, projection effects, and gas temperature influence the inferred CGM turbulence driving scale. We find that projection significantly lowers the VSF normalization but we do not find significant differences in the slope between 3D VSFs and emission-weighted projected 2D VSFs. We find that the size of the area used to measure the VSF, which can be thought of as the size of the emission nebula for a given instrument sensitivity, correlates directly with the turnover location in the VSF. These dependencies should be considered when using VSFs to interpret CGM turbulence from emission data, as projection, resolution and sensitivity constraints, and the temperature of the gas probed will all have a measurable effect on the VSF structure and the corresponding inferred turbulent properties. 
\end{abstract}


\section{Introduction} \label{sec:intro}

Galaxies are surrounded by an extended, gaseous environment commonly referred to as the circumgalactic medium (CGM). The CGM encompasses all gaseous surroundings of a galaxy out to the virial radius of its dark matter halo \citep{Tumlinson2017}, which includes galactic outflows from the central galaxy and any satellites \citep{Thompson2024}, gaseous inflows arriving from the cosmic web or intergalactic transfer from other galaxies \citep{Keres2005,Dekel2009,AnglesAlcazar2017}, and any other gas that is circulating, turbulent, or in hydrostatic equilibrium within the halo. Observations and simulations alike have uncovered the multiphase nature of the CGM, finding gas at a variety of temperatures detectable at wavelengths from radio to X-ray \citep{Putman2012,Tumlinson2017,FaucherGiguere2023,Crain2023}.

Unraveling the gas dynamics occurring within the CGM provides insight on how galaxies obtain the fresh fuel they need for new star formation and how galactic outflows enrich the intergalactic medium with metals. Inflows, outflows, and other gas dynamics interact to create a turbulent medium within the CGM. Turbulence can have a significant impact on the state of the CGM: it can provide pressure support \citep{Oppenheimer2018,Ji2020,Lochhaas2020,Lochhaas2023}, its dissipation can be a source of heating \citep{Zhuravleva2014}, and it can generate density fluctuations that enhance cooling and promote condensation of cool gas \citep{Voit2018,Fielding2020,Gronke2022,Wibking2025}. There are many possible driving sources of turbulence in the CGM, including outflows launched by stellar feedback or active galactic nuclei (AGN) \citep{Fielding2017,Li2020,Wang2021}, galaxy mergers \citep{Sparre2022,Bennett2022}, motion of satellite galaxies along their orbits \citep{Kim2007,Ruszkowski2011}, and gas accretion from outside the halo \citep{Vazza2009,Robertson2012,Goldner2025}. Disentangling the relative contributions to CGM turbulence from each source is non-trivial because turbulence is most easily measured and described statistically, so a significant amount of data is required.

One of the most common ways to observe the CGM of $L^\star$ galaxies is through rest-ultraviolet (UV) absorption lines in the spectra of background light sources, such as bright quasars \citep{Tumlinson2011,Werk2014,Chen2020}. Through detailed ionization modeling of absorbers, the density and temperature of the CGM of a foreground galaxy can be inferred, and thus the non-thermal broadening from turbulence \citep{Rudie2019,Chen2023a}. However, absorption line studies of the CGM are typically limited by only 1--2 spectra for any given foreground galaxy, so it is difficult to obtain enough absorption data for a single system to measure the turbulence therein. This is typically limited to a handful of systems where either lensed quasars \citep{Rauch2001,Augustin2021} or lensed arc tomography \citep{Tejos2021,Shaban2025} of a background galaxy can be used to obtain multiple lines of sight (LOS) through a single galaxy's CGM \citep[or for our closest neighbor M31, whose CGM is broad enough on the sky for many background sources,][]{Lehner2020}. Statistical turbulence measurements in absorption therefore require combining data across multiple galaxy systems, which inherently assumes similar turbulence statistics in all systems, which may not be the case.

With the advent of integral field unit (IFU) spectrographs, such as the Keck Cosmic Web Imager \citep[KCWI][]{Morrissey2018} on the Keck telescopes and the Multi-Unit Spectrographic Explorer \citep[MUSE][]{Bacon2010} on the Very Large Telescope (VLT), the possibility now exists to obtain kinematic (and spatial) information from a \emph{single} galaxy's CGM in emission. Emission from the CGM is expected to be inherently faint at low redshifts due to low gas densities \citep{Saeedzadeh2025}, but a handful of galaxy systems at $z\sim$ 0--1 have been observed to have extended [\ion{O}{2}] and [\ion{O}{3}] halos that can be kinematically mapped using LOS velocity of the emission line in each spaxel \citep{Chen2024,Nielsen2024}. With 2D spatial and 1D kinematic information, turbulent statistics can be measured through the use of a velocity structure function (VSF).

The VSF is the real-space equivalent of the kinetic energy power spectrum, and measures the variation of velocity at many scales. With a limited number of spaxels, it is often easier to compute a VSF than the power spectrum directly. Observationally, VSFs have been computed in star forming regions of the ISM \citep{ODell1987,Arthur2016,Xu2020,Ha2022}, from cool gas H$\alpha$ filaments in galaxy clusters \citep{Li2020}, the tails of jellyfish galaxies in galaxy clusters \citep{Li2023,Ganguly2023}, and [\ion{O}{2}] and [\ion{O}{3}] emission halos of $z\sim$ 0--1 quasar host galaxies \citep{Chen2023b,Chen2024}. The VSF encodes information about the turbulent cascade within the gas from which it is measured: the slope of the VSF, typically a power law, is directly related to the rate of energy transfer from large-scale turbulent eddies to smaller eddies. The flattening or turnover location of the VSF on large scales is often used to infer the scale of the turbulent injection that begins the cascade, and the small-scale end can provide information on the turbulence dissipation mechanisms. Thus, measuring the VSF from an emission halo observed with an IFU may be a direct way to probe the properties of turbulence in the CGM of the galaxies hosting the emission.

However, IFU observations of emission halos obtain information in just two spatial directions (projected on the plane of the sky) and one velocity direction (along the line of sight). A particular emission line will originate from gas of a specific phase, and emissivity is strongly dependent on the gas density. In addition, observational effects like spectral and spatial resolution, seeing, and sensitivity of the instrument will affect the quality and quantity of the data from which the VSF is measured. Constraints that affect the measured VSF strongly affect the inferred turbulence statistics. Observing the full 3D spatial and velocity structure of a distant galaxy's CGM is not possible, but we can use simulations, where such information \emph{is} available, to forward-model into observational space. By comparing the ``true" underlying turbulence in full 3D in the simulation to a mock observation made from the same simulation, we can determine the effects on the VSF of projection, reducing the velocity information to one dimension, spatial resolution, size of the emission halo, and observing gas within a narrow temperature range.

In this paper, we use the Figuring Out Gas \& Galaxies In Enzo (FOGGIE) simulations to measure VSFs in both 3D and from mock ``observations" to explore the impact of various observational constraints. In Section~\ref{sec:methods}, we give an overview of FOGGIE (\S\ref{subsec:FOGGIE}), describe how we measure the VSF from the full 3D simulation information (\S\ref{subsec:VSFs}), and explain how we remove the disk of the galaxy to focus on turbulence in the CGM (\S\ref{subsec:disk_removal}). Section~\ref{sec:3D_VSFs} presents the impact of changing the pixel resolution or measurement area size on the 3D VSFs (\S\ref{subsec:3D_res_box}) and how the 3D VSF varies with galactocentric radius (\S\ref{subsec:radial}). Section~\ref{sec:2D_VSFs} moves on to the 2D projected VSFs, where we describe how we create mock velocity maps and measure the VSF (\S\ref{subsec:projections}), compare 3D and 2D projected VSFs (\S\ref{subsec:proj_effects}), again explore the impact of pixel resolution and measurement area size (\S\ref{subsec:2D_VSF_res_box}), and finally explore the impact of limiting the VSF measurement to a narrow range in temperature to mimic measuring a VSF from a single observed emission line (\S\ref{subsec:2D_VSF_temp}). Section~\ref{sec:discussion} discusses some caveats (\S\ref{subsec:caveats}) and compares our findings to other observed (\S\ref{subsec:obs_compare}) and simulated (\S\ref{subsec:sim_compare}) turbulence studies, and we summarize and conclude in Section~\ref{sec:summary}.

\section{Methods} \label{sec:methods}


\subsection{Figuring Out Gas \& Galaxies In Enzo (FOGGIE)}\label{subsec:FOGGIE}

The Figuring Out Gas \& Galaxies In Enzo (FOGGIE) simulations are a suite of cosmological zoom-in simulations using the code Enzo \citep{Bryan2014,BrummelSmith2019}. The full details of the simulation setup can be found in \citet{Peeples2019,Simons2020,Wright2024}, so here we give a brief overview of the components relevant for this work.

FOGGIE focuses on six galaxies selected to be roughly Milky Way-mass (halo mass of $\sim10^{12}\,\mathrm{M}_\odot$) by $z = 0$, with no major mergers (greater than $1:10$ mass ratio) at redshifts less than $z = 2$. Stars form using the scheme of \citet{Cen1992} and supernova explosions inject thermal energy (at the rate of $10^{51}$ ergs per $100\,\mathrm{M}_\odot$ of stellar mass formed) into their immediate surroundings.

Enzo uses adaptive mesh refinement (AMR) to increase the resolution of its cubic simulation cells where necessary. Cells are refined to the next level when their density becomes large or their cooling time becomes small. In addition, surrounding each zoom-in galaxy is a box with side length 288 comoving kpc (ckpc), called the forced refinement region, wherein the resolution is not allowed to de-refine to cell sizes larger than 1.1 ckpc on a side \citep[see][for a more detailed description of the refinement criteria used in FOGGIE]{Wright2024}. The forced refinement region allows for the gas in the CGM, far from the galaxy, to be simulated at high resolution. Both previous FOGGIE papers and other studies with enhanced CGM resolution have shown that resolving the CGM reveals more structure in the cooler gas phases \citep{Peeples2019,Hummels2019,vandeVoort2019,Bennett2020,Ramesh2024} and, of particular importance for this work, more kinematic structure in all gas phases, including the warmer phases \citep{Lochhaas2021,Lochhaas2023}. Note that we deposit simulation cell information on uniform grids before calculating the VSF, so the AMR structure does not complicate the calculation of the velocity structure function (see Section~\ref{subsec:VSFs}).

For this work, we focus on Maelstrom, which is one of the FOGGIE simulated galaxies, at $z = 0$. Maelstrom has a virial mass of $10^{12}\,\mathrm{M}_\odot$ and a virial radius of 212 kpc, a fairly quiescent star formation rate of $2\,\mathrm{M}_\odot$ yr$^{-1}$, and no active galactic nucleus, making it similar to the Milky Way. We note that the observations from which CGM velocity structure functions are typically measured \citep[e.g.,][]{Chen2023b,Chen2024} focus on group galaxies that host quasars at $z\approx0.5$--1, since these systems are the few that are bright enough (due to ionization from the quasar) to have detectable emission from their CGM. While these systems are physically different from the simulated FOGGIE galaxy we investigate here, our focus is comparing 3D turbulence to the 2D velocity structure functions derived from these kinds of observations to determine how observational biases may affect the turbulent properties inferred. As such, we do not make any direct comparisons with observed structure functions. Throughout this paper, we use only the $z=0$ snapshot of Maelstrom, but verified we find similar relations between 3D and 2D VSFs in Maelstrom's $z=1$ snapshot and in one another simulated FOGGIE galaxy at $z=0$.

\subsection{Velocity Structure Functions } \label{subsec:VSFs}

Second-order velocity structure functions (VSFs) are a common method of characterizing turbulence in gas, which we here apply to understand turbulence in the CGM \citep[e.g.,][]{Pope2000}. The VSF is defined as 
\begin{equation}
S_2(s) = \langle | \mathbf{v}(\mathbf{x}) - \mathbf{v}(\mathbf{x}+\mathbf{s}) |^2 \rangle
\end{equation}
where $\mathbf{v}(\mathbf{x})$ and $\mathbf{v}(\mathbf{x}+\mathbf{s})$ are the 3D velocity vectors at two points $\mathbf{x}$ and $\mathbf{x}+\mathbf{s}$ where $s$ is the vector separation between the two locations. The angle brackets indicate an average over all pairs of points with similar separations $|\mathbf{s}|$. Note that $S_2$ is the ``total" quantity of the second-order VSF, where the velocity difference $\mathbf{v}(\mathbf{x}) - \mathbf{v}(\mathbf{x}+\mathbf{s})$ is a vector quantity with no preferred direction relative to $\mathbf{s}$.

The slope of the VSF provides information about the type of turbulence in the medium and the ``flattening" or ``turnover" of the VSF is expected to occur at the physical scale where turbulence is driven, the injection scale where the turbulent energy cascade begins. Often, this scale is tied to other physical scales in a system to determine the process responsible for driving turbulence \citep{Li2020}. For Kolomogorov turbulence, which is isotropic, subsonic, and incompressible, it is expected that the slope of the second-order VSF scales as $S_2(s) \propto s^{2/3}$ \citep{Kolmogorov1941}. The overall normalization of the VSF is connected to the strength of turbulent velocities: faster turbulence, which produces a larger velocity dispersion, increases the normalization.

In practice, to calculate the second-order VSF, we must select a large number of pairs of spatial locations from the simulation, calculate the difference in 3D velocity between each point in the pair, then bin each pair in 3D spatial separation and average among all pairs in each bin. We start by selecting a subset of the simulation volume that is a cube 200 kpc on a side and is centered on the galaxy, although we explore the impact of picking different box sizes in \S\ref{subsec:3D_res_box}. Because the simulation is allowed to refine and de-refine cells at runtime according to its refinement criteria, the selected volume does not necessarily represent a grid of uniform resolution. To correct for this, we deposit simulation cells onto a uniform grid by averaging the properties in cells at a higher resolution than the grid. We do not use any region of the simulation with worse simulation resolution than the uniform grid resolutions we choose, so we only down-sample, not up-sample. We evaluate the effect of choosing different resolutions for this grid in \S\ref{subsec:3D_res_box} and pick 2.2 kpc as our fiducial resolution (a factor of two down-sampled from the forced refinement region in FOGGIE). This gives us a cubic volume that is made up of 91$^3$ gas cells. We select pairs of cells randomly from this volume because it is computationally infeasible to calculate the VSF from every pair of cells. We use 6000 pairs of randomly-selected cells within this volume, which samples $\sim 0.8\% $ of the volume and represents a balance between computational feasibility and accuracy\footnote{Among four trials of different sets of 6000 point pairs, we find the computed VSF does not shift by more than 5\% in its normalization or shape.}. We bin each pair of cells by the separation between the two points in the pair, using separation bins from 0 to 210 kpc in steps of bin size 2 kpc.

\subsection{Galaxy Disk Removal} \label{subsec:disk_removal}
To most purely analyze the turbulence solely from the CGM, we want to remove the galaxy disk. To do this, we create a density cut, defined as $2 \times 10^{-27}$ g cm$^{-3}$, and remove from consideration all gas with a density higher than this value, leaving just the CGM behind \citep[as done in][]{Lochhaas2023}. We only select pairs of cells that fall in this low-density region.

\begin{figure*}
  \centering
  \includegraphics[width=0.7\linewidth]{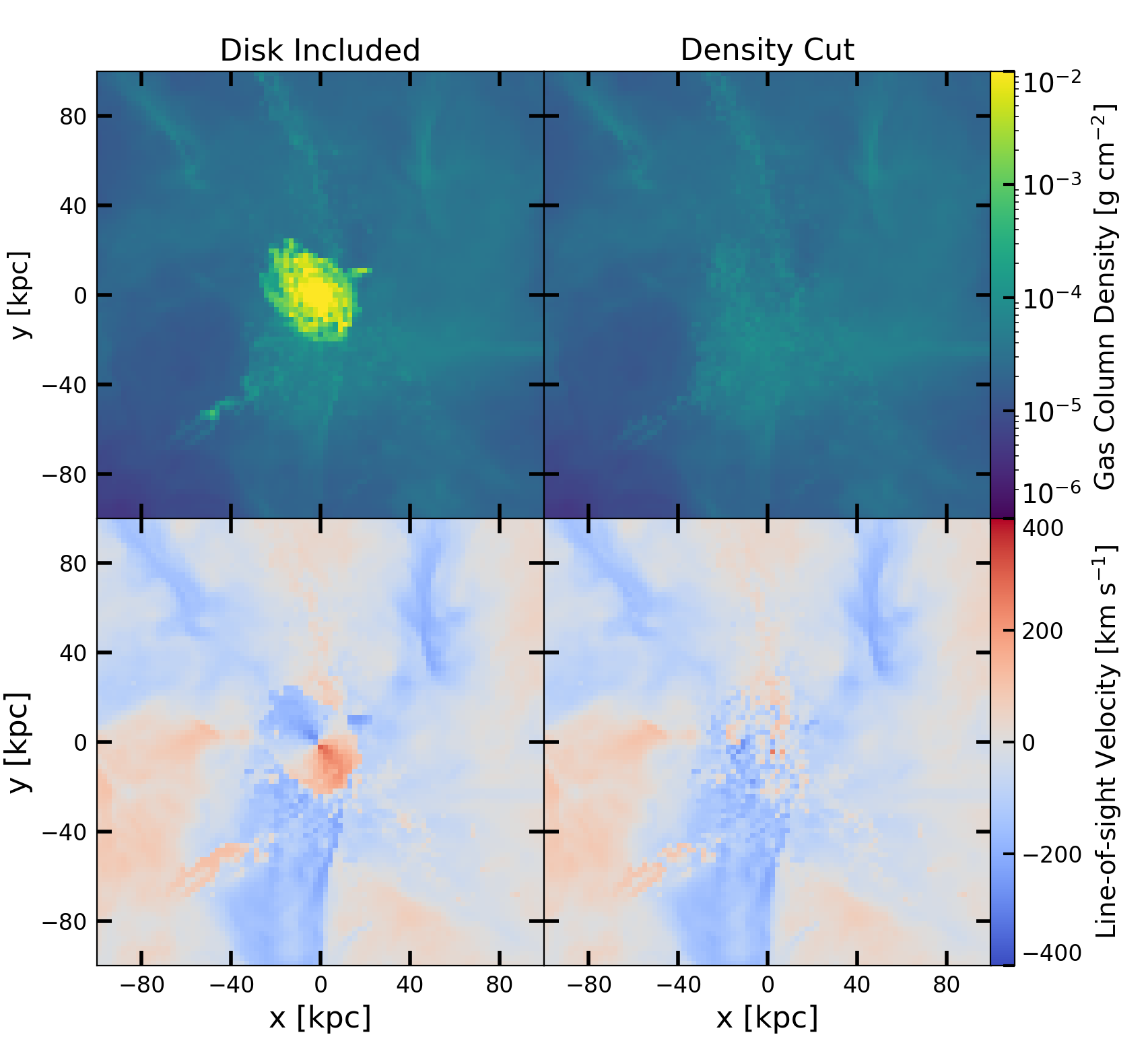}
  \caption{\emph{Top panels:} Density projections of the Maelstrom simulated galaxy at $z=0$, both with (left) and without (right) the central galaxy's gas disk. \emph{Bottom panels:} Line-of-sight velocity, weighted by the square of the gas density to mimic emissivity (see \S\ref{subsec:projections}), of all gas (left) and with the disk removed by density cut before projection (right).}
  \label{fig:diskprojs}
\end{figure*}

Figure~\ref{fig:diskprojs} shows projections of the gas density (top row) and mean LOS velocity (bottom row) for the Maelstrom simulated galaxy at $z = 0$. The left-hand panels show all gas in and surrounding the galaxy, while the right-hand panels show the gas that is left behind when the disk is removed using a density cut. The galaxy disk, viewed from this random inclined angle, shows LOS velocities between $\pm \sim\!150$ km/s with a clear gradient from top-left to bottom-right. Removing the disk eliminates the central velocity gradient, leaving behind only the LOS velocities of the CGM gas.

\section{Properties of 3D VSFs} \label{sec:3D_VSFs}

\subsection{Effects of Resolution and Box Size} \label{subsec:3D_res_box}
We begin by examining the effect of different grid resolutions on the structure of the VSF by depositing the same simulation data onto uniform grids of different resolutions. We do \emph{not} re-run the simulation at different resolutions, which would likely have a significant effect on the evolution of the turbulence. Instead, our method can be thought of as analogous to ``observing" the same underlying kinematic structure at different observational resolutions.  We compute the VSFs at resolutions of 1.1, 2.2, and 4.4 kpc.

\begin{figure}
  \centering
  \includegraphics[width=\linewidth]{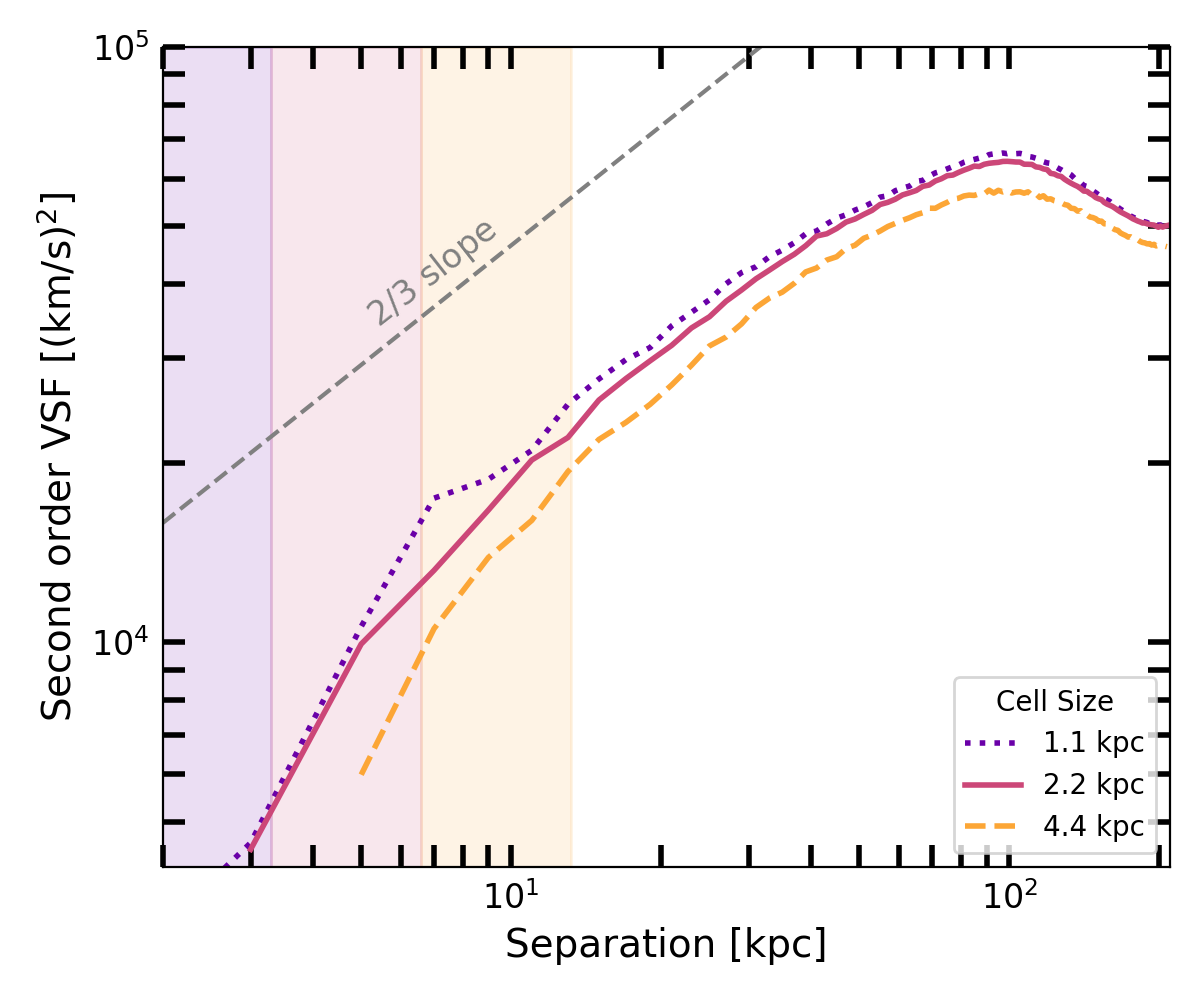}
  \caption{The second-order VSFs calculated from a box of size 200 kpc on a side, centered on the galaxy, with the disk removed using a density cut. The three curves represent the VSF calculated on uniform grids of different cell sizes: 4.4 kpc (yellow dashed), 2.2 kpc (pink solid, fiducial resolution), and 1.1 kpc (purple dotted, the resolution of the forced refinement region in FOGGIE). Colored shading of the same color as the curves show pair separations that are within $3\times$ the cell size, indicating regions where numerical resolution effects bias the VSF calculation. The gray dashed line demonstrates the expected scaling from Kolmogorov turbulence. Changing the simulation resolution does not strongly affect the resultant VSFs, however the normalization is slightly higher at higher resolutions.}
  \label{fig:3D_resolution}
\end{figure}

Figure~\ref{fig:3D_resolution} shows the second-order VSFs calculated from our fiducial box at different resolutions of the uniform grid. At higher resolutions, the normalization of the VSF is higher. This is likely because at lower resolution, the velocity of each uniform gridcell is smoothed over neighbouring cells, thus reducing the measured velocity dispersion. However, despite this difference, the slope of each line is consistent with $\sim2/3$, matching Kolmogorov turbulence, and the turnover location remains at $\sim100$ kpc across resolutions. This indicates that the grid resolution on which the VSF is calculated does not much affect the turbulence properties inferred from the VSF. We further discuss this in Section~\ref{subsec:2D_VSF_res_box} in the context of beam-smearing for different degrees of seeing.

\begin{figure}
  \centering
  \includegraphics[width=\linewidth]{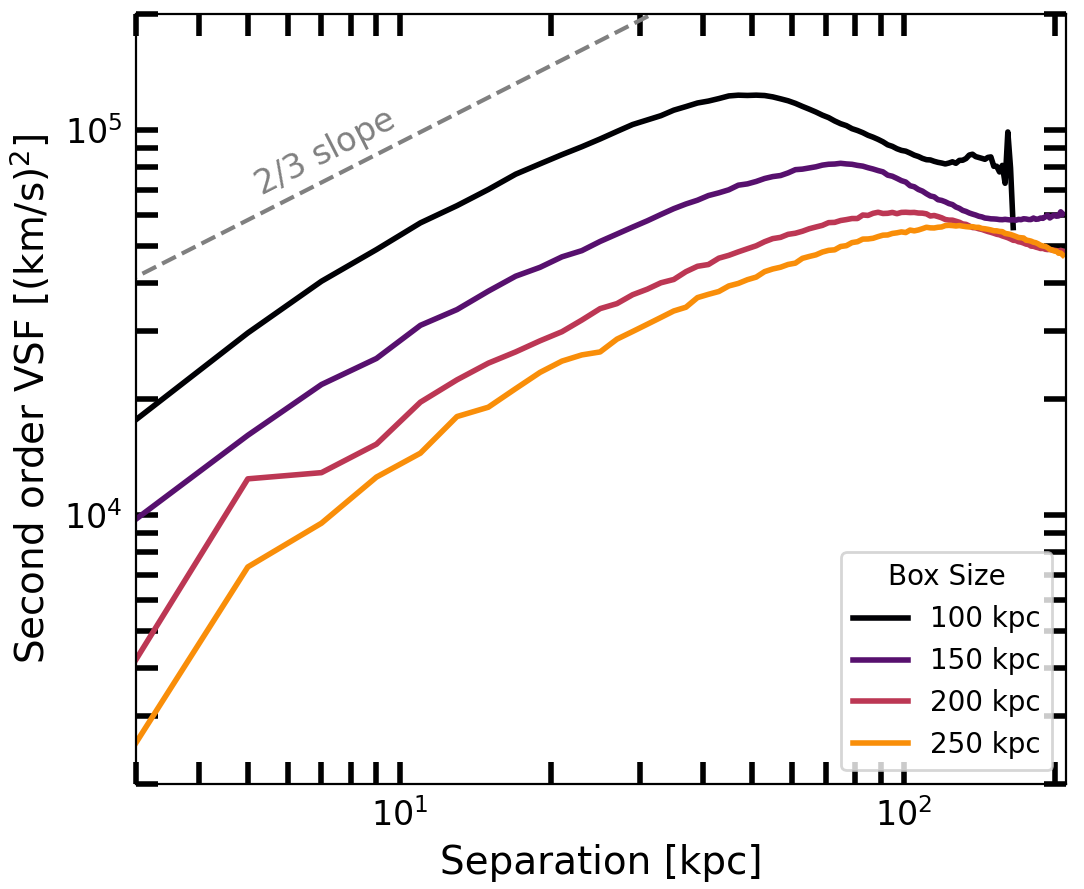}
  \caption{The second-order VSFs calculated from boxes of different sizes, each centered on the galaxy and each calculated at a grid resolution of 2.2 kpc. The black curve represents the smallest box of 100 kpc, with curves of increasing lightness representing larger boxes up to 250 kpc (light orange). The gray dashed line indicates the expected $2/3$ slope of Kolmogorov turbulence. When calculated over a smaller box, the separation at which the turnover of the VSF occurs decreases while the overall normalization increases.}
  \label{fig:3D_box_size}
\end{figure}

Figure~\ref{fig:3D_box_size} shows the effect of measuring the VSF with differently sized boxes as the measurement area. Each box is centered in the same place, on the galaxy center, and each uses the same resolution for the uniform grid of 2.2 kpc. In contrast to changing resolution, changing the box size of the volume used to calculate the VSF strongly affects both the normalization and the turnover location of the VSF. This indicates that the flattening or turnover location of the VSF may not be a simple one-to-one mapping to the physical driving scale of turbulence. This interpretation is supported by Appendix~\ref{appdx:box_size}, wherein we find that the turnover location of the VSF does \emph{not} correlate with the driving scale, but rather with the size of the measurement region, if the true driving scale of the turbulence is larger than that size. Other works have similarly found an effect of measurement region size on the VSF turnover or flattening point \citep{GarciaVazquez2023,Ha2022,Chen2024}. We explore this more in the next subsection, and in Section~\ref{subsec:2D_VSF_res_box}.

\subsection{VSFs at Different Galactocentric Radii} \label{subsec:radial}
To attempt to separate measurement region size with true variations in the turbulence driving scale, we split the simulated CGM into radial shells and calculated the VSF within each shell. We used five radial bins of width 20 kpc between 0 and 100 kpc from the center of the galaxy.

\begin{figure}
  \centering
  \includegraphics[width=\linewidth]{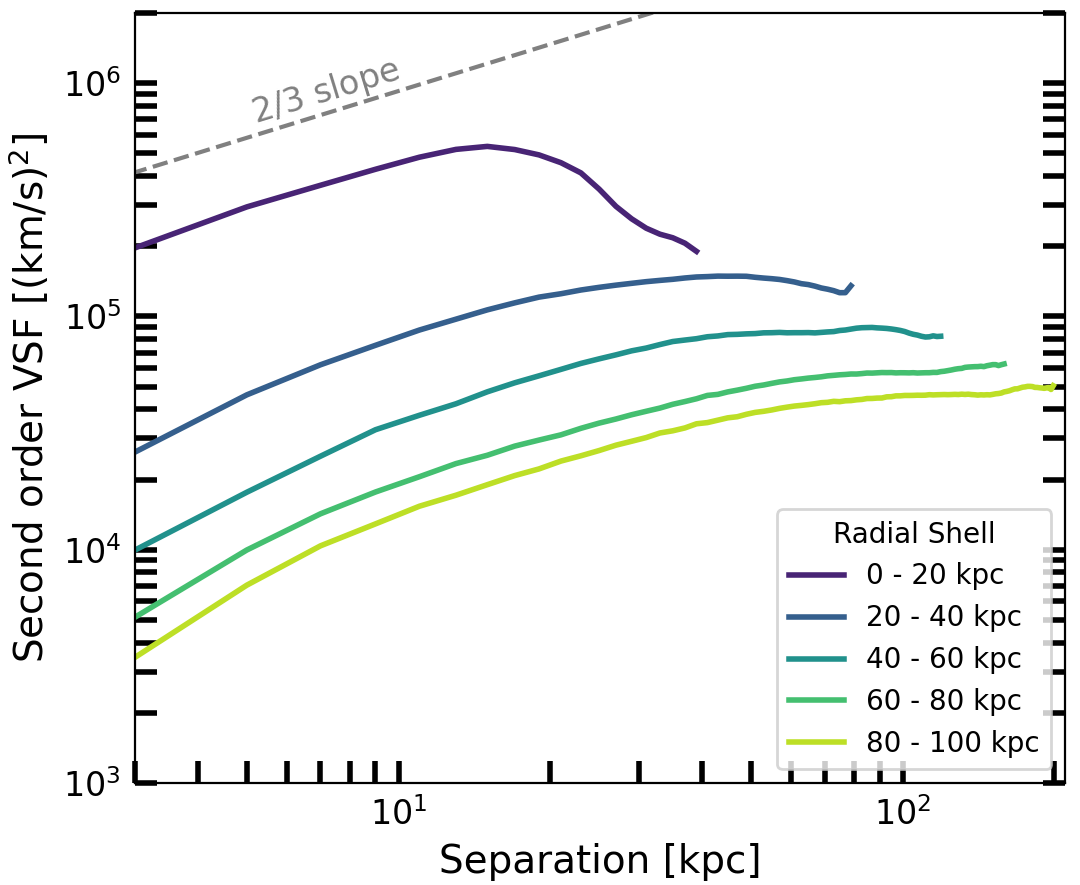}
  \caption{Second order VSFs calculated within radial shells of width 20 kpc increasing in distance from the galaxy. Colors of curves lighten from dark blue to light green as distance from galaxy increases, as indicated in the legend. The Kolmogorov turbulence expected slope of 2/3 is represented in the light gray dashed line. Although the normalization decreases and turnover location increases with increasing radius, the slopes are consistent with Kolmogorov turbulence.}
  \label{fig:radial_bins}
\end{figure}

Figure~\ref{fig:radial_bins} shows the VSFs calculated within each radial shell. The VSFs in smaller radial shells have higher normalizations than at larger radii and also exhibit smaller turnover locations. The gradual increase of turnover location indicates either that the turbulence of the CGM is being driven at larger scales as the distance from the galaxy increases, or that the smaller radial shells are smaller than the driving scale of the turbulence in that region (see Appendix~\ref{appdx:box_size}), as both effects would produce a similar shift of the flattening scale to larger values. In the largest shells, the flattening location appears to approach a similar value, indicating that perhaps shells $\gtrsim50$ kpc are correctly capturing the driving scale of the turbulence.

To test this explicitly, we also measured the VSF in boxes of side-length 50 kpc centered on different locations in the galaxy halo, stepping outward from the center of the galaxy in a random direction in steps of 20 kpc. Since all the boxes are the same size, this experiment limits the dependence on measurement region size to focus purely on the driving scale of the turbulence. We found that the turnover point of the VSFs measured from these boxes shifted to larger scales as the box moved further from the galaxy, up to $\sim40$ kpc, and beyond this distance no obvious flattening or turnover point occurred within the 50 kpc box. This trend suggests that the driving scale of the turbulence is truly smaller closer to the galaxy, but increases to larger scales moving to further distances to the galaxy, up to the point where the 50 kpc size of the measurement box used can no longer accurately capture the turnover point because the driving scale has shifted to scales much larger than this.

On the other hand, the decrease in normalization with increasing radius cannot be explained purely by changing the measurement region size, so is likely indicating the strength of turbulence is truly declining with distance from the galaxy. This suggests that the larger box sizes are dominated by the turbulence further from the galaxy, since increasing the box size as in Figure~\ref{fig:3D_box_size} produces the same gradual shifting effect to lower normalizations. The same effect is seen with idealized turbulent velocity fields generated with known strength and turnover location that vary with distance from the center of the halo in Appendix~\ref{appdx:many_driving}, providing strong support for this interpretation.

Determining the source(s) of turbulent driving in a galaxy halo is non-trivial, but the increasing driving scale with increasing distance from the galaxy (if this result is not due to too-small measurement regions) suggests that the turbulence driver at the outskirts of the halo is a larger-scale process than that in the inner regions of the CGM. \citet{Goldner2025} showed that spherical accretion onto and through a galaxy halo, as part of a cooling flow, can enhance any existing velocity perturbations through adiabatic compression. This both increases the turbulent velocity toward the center, such that it scales as $\propto r^{-1}$, and decreases the turnover location of the VSF, such that the turnover location scales as $\propto r$. We find a very similar evolution of the VSF as a function of radius in Figure~\ref{fig:radial_bins}, so it is possible that much of this evolution is driven by accretion of gas onto the halo. However, we note that \citet{Goldner2025} derived accretion-driven turbulence in idealized simulations without feedback, while the FOGGIE simulations do have feedback, as well as mergers, which have also been shown to drive turbulence in both galaxy halos \citep{Iapichino2013,Sparre2022} and clusters \citep{Bennett2022}. Mergers and feedback both stir turbulence closer to the center of the halo, so the FOGGIE galaxies may show enhancement in turbulence in the inner regions over that expected solely from the compression of the accretion flow. \citet{Lochhaas2023} indeed found much stronger turbulent support in the inner regions of the FOGGIE galaxies' CGM than in the outskirts.

\section{2D Projected VSFs} \label{sec:2D_VSFs}
\subsection{Making Velocity Maps to Mimic Observations} \label{subsec:projections}

In observations of the CGM, full 3D information about positions and velocities of gas are not known. Instead, projected VSFs are computed from IFU datacubes using the emission line centroids as a LOS velocity and projected 2D separations between each pixel where emission is measured. To mimic the observations, we project the 3D simulation into 2D by choosing one of the Cartesian axes of the simulation (which are random with respect to the simulated galaxy orientation) and finding, for each pixel, the density-squared weighted mean of the line-of-sight velocity along the full depth of the cube (100 kpc). The density-squared weighting mimics the scaling of emission with gas density, and by using the mean of the line-of-sight velocity, we mimic a velocity centroid for the emission\footnote{If an IFU spaxel exhibits multiple emission components, a choice must be made for how to compute the LOS velocity. The velocity could be taken from only the strongest component, ignoring all others, or it could be taken from a single-component fit even if a better fit could be achieved with more components, or it could be the flux-weighted average velocity. \citet{Chen2023b} shows these choices do not strongly affect the shape of the VSF, but may alter its normalization somewhat. The method we choose here, the emission-weighted average LOS velocity, is most similar to the flux-weighted average velocity method, but becomes equivalent to a centroid velocity if there is only a single component. \citet{Esquivel2005} find the velocity centroids to recover turbulent statistics well for subsonic turbulence.}. A complete forward-modeling of the simulation into expected emission lines is an avenue for future research but is beyond the scope of this work.

In 2D, the distance between the two selected points is reduced to a projected distance, and the velocity difference is not a vector difference but simply the difference between line-of-sight velocities in each pixel. In 2D, there are fewer spatial pixels in the image than in the full 3D datacube, so it is computationally feasible to compute the VSF from every pair of pixels in the image, unlike in 3D. Throughout this paper, we use random pairs of cells to calculate the VSF in 3D and use all pairs of cells when calculating the VSF from the 2D projections.

\begin{figure*}
  \centering
  \includegraphics[width=\linewidth]{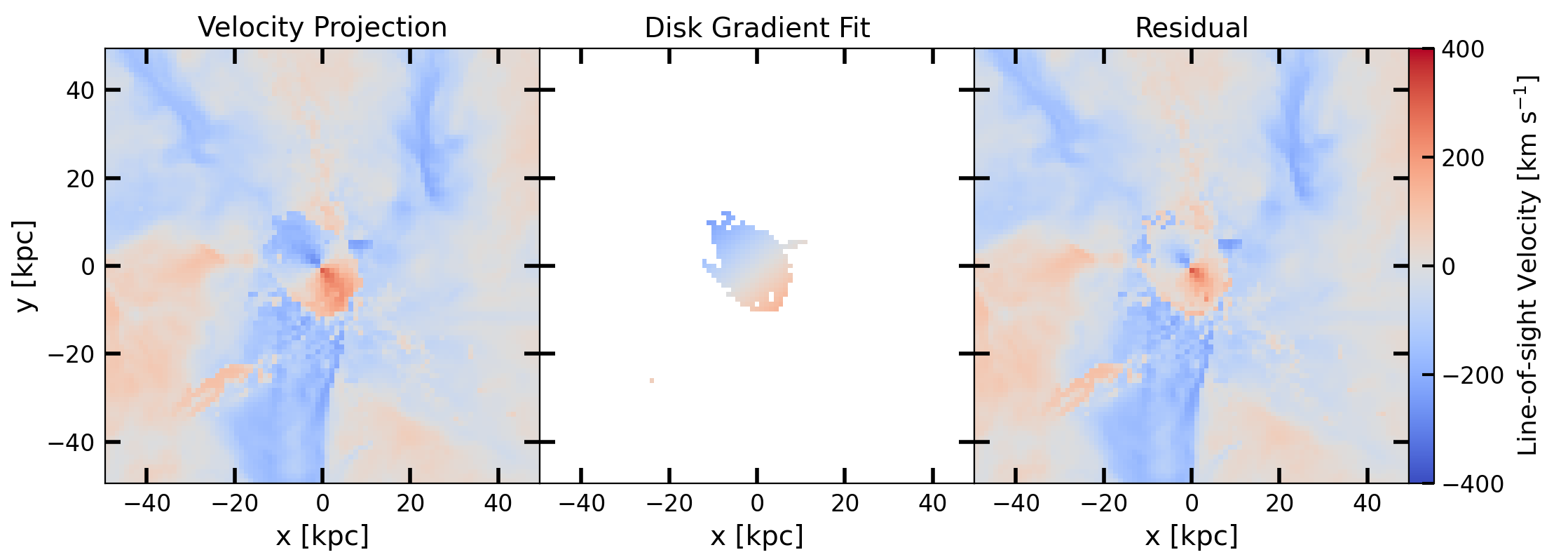}
  \caption{\emph{Left:} Density-squared-weighted mean LOS velocity of all gas of Maelstrom simulated galaxy. \emph{Center:} Isolated 2D velocity gradient fit of the disk created by separating the disk using a density cut and calculating the best fit gradient. \emph{Right:} Residual image created by subtracting the disk gradient fit from the velocity projection. This residual is used as a way to replicate observational methods of disk removal when isolating the CGM.}
  \label{fig:velgrad}
\end{figure*}

While simulations simplify masking out a certain density range to remove the galaxy disk, allowing for an isolated analysis of the CGM, observers may not have the luxury of this information. Instead, it is common to first remove any velocities that may indicate bulk disk rotation by fitting and subtracting off a velocity gradient before calculating the VSF \citep{Chen2023b}. We perform this same process to most closely match observational VSFs and show the steps of the process in the panels of Figure~\ref{fig:velgrad}. We start by isolating the region of the line-of-sight velocity image (left panel of Figure~\ref{fig:velgrad}) by selecting cells with surface gas density above 10$^{-3.5}$ g/cm$^{-2}$, corresponding to the disk as determined by-eye from the top left panel of Figure~\ref{fig:diskprojs}. Then, we fit a linear velocity gradient of the form $v(x, y) = ax + by + c $ to the center of the disk region (center panel of Figure~\ref{fig:velgrad}) and subtract the gradient off from the region identified to be disk-like (right panel of Figure~\ref{fig:velgrad}), leaving the velocity field outside the disk untouched. We then calculated the 2D projected VSF from the residual velocity field. The residual shows that a gradient fit is not perfect, because the disk velocities are much higher near the center than further out in the disk. However, the area of the image with such large rotation velocities is very small, so is unlikely to strongly bias the VSF (and indeed we find it does not in Section~\ref{subsec:proj_effects} below).

\subsection{Projection effects on the VSF}\label{subsec:proj_effects}

\begin{figure}
    \centering
    \includegraphics[width=\linewidth]{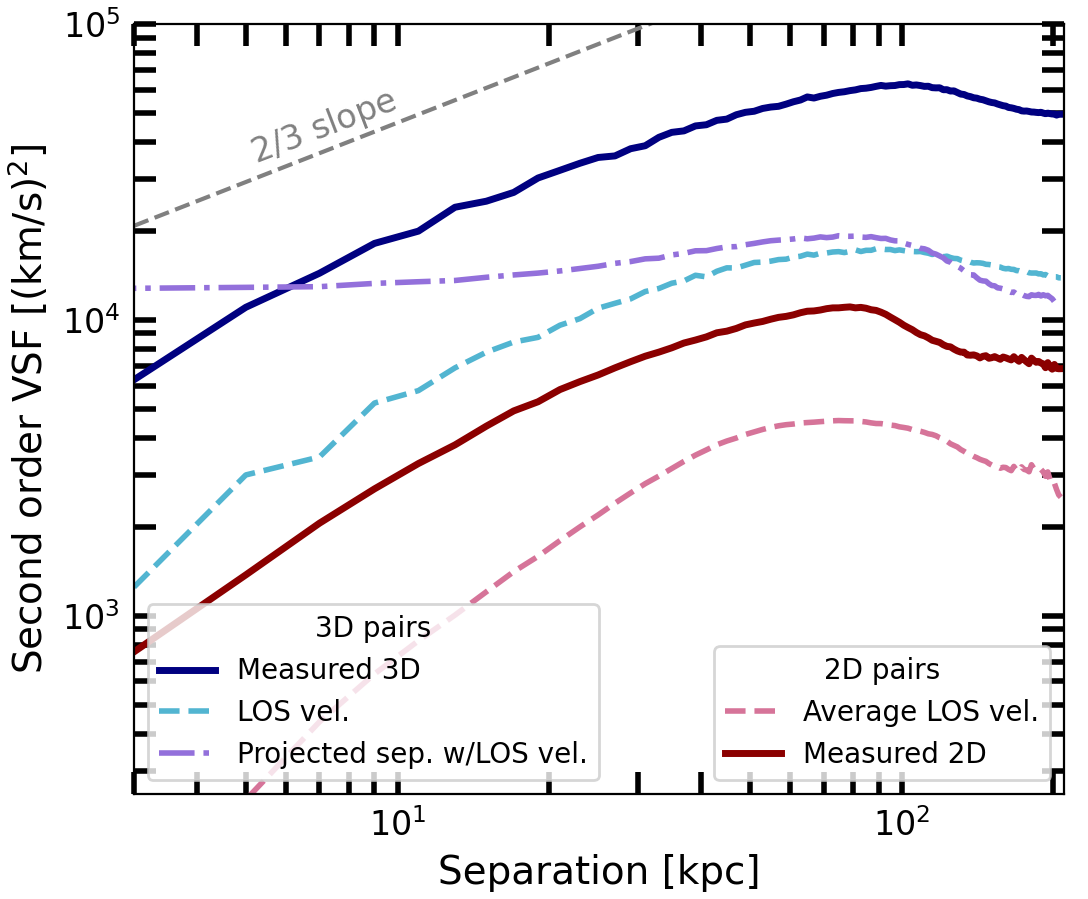}
    \caption{The effects of different rungs of the projection ladder on the VSF (see text). The solid dark blue and red lines represents the VSF measured from all the gas in the simulation (including the central galaxy disk), measured using 3D and 2D separations respectively. The light blue dashed curve represents the VSF calculated using 3D separations but only the line of sight direction of the velocity (rung 1). This line is lower in normalization than the full 3D VSF by a factor of 3. The purple dash-dotted curve represents the VSF calculated using 2D projected separations and LOS velocity (rung 2). The pink dashed line represents the 2D VSF calculated using the strict average LOS velocity rather than the density-squared-weighted mean (rung 3). The 2/3 Kolmogorov scaling law slope is represented in the gray dashed line.}
    \label{fig:projection_ladder}
\end{figure}

In moving from a 3D VSF to a 2D projected VSF, there are both geometric effects and physical effects of the projection that contribute to the difference between the VSFs. We show a ``ladder" of steps to move from 3D to 2D in Figure~\ref{fig:projection_ladder} to highlight which features are related to geometry and which are related to the particular physics of making a mock emission-weighted velocity projection from our simulation. In Appendix~\ref{appdx:projection_ladder}, we show a similar ladder for the idealized case of an isotropic turbulent velocity field. 

The first rung of the ladder is to reduce the velocity dimensionality from a 3D vector to only the 1D LOS velocity. The light blue dashed curve shows the VSF calculated using 3D vectors for the spatial separation between pairs of points, but using only the LOS direction of the velocity for the velocity difference. This VSF is similar in shape and slope to the fully 3D VSF (dark blue solid curve), but lower in normalization by a factor of 3, because it uses 1/3 of the velocity dimensions. Note that this does \emph{not} represent the decomposition of the VSF into transverse ($D_{NN}$) and longitudinal ($D_{LL}$) directions \citep[e.g.,][]{Pope2000}, where, for Kolmogorov turbulence, it is expected that $D_{NN}=\frac{4}{3}D_{LL}$. Because we do not restrict the point-pair selection to be purely in the plane perpendicular to the line of sight, we are not measuring exactly $D_{NN}$.

The second rung is to calculate the VSF using 2D separations (in the plane of the sky) and 1D LOS velocity. We still select pairs of points randomly, so two points may have very different depths along the line of sight but have a small projected separation on the sky. This VSF is shown in Figure~\ref{fig:projection_ladder} as the purple dash-dotted curve, and this VSF is significantly shallower in slope and approaches a flat value at small separations. This is because we have essentially erased much of the pair separation information and are mixing points that may be separated by quite large values (along the line of sight) into small (projected) separation bins. Appendix~\ref{appdx:projection_ladder} shows how the depth of the gas region probed affects this slope: generally, the deeper the region probed, the shallower the VSF becomes. Physically, this scenario aligns with a picture where the observed emission line originates from a single, localized gas cloud somewhere along the line of sight. This is unlikely to be the true picture in a CGM emission halo because such localized, sparse clouds would produce a much ``patchier" emission halo than is typically observed.

Next, for the third rung of the ladder, we compute the VSF using 2D projected separations in the plane of the sky and averaging the LOS velocity everywhere along the line of sight. Note that we do not yet use the density-squared weighting for producing velocity maps from mock emission, as shown in Figure~\ref{fig:velgrad}. The simulation data is deposited on a uniform grid before calculation (see \S\ref{subsec:VSFs}), so a strict average along the LOS is a volume average. This VSF is shown in Figure~\ref{fig:projection_ladder} as the  pink dashed curve. It is lower normalization than the first rung and also shows a slightly steeper slope. Because the velocity has been averaged across the full depth of the line of sight, power has been lost both overall and particularly at smaller scales, leading to both the overall normalization and steepening below the injection scale. Appendix~\ref{appdx:projection_ladder} shows how both the steepening and normalization are dependent on the depth of the region probed: the deeper the region, the steeper the slope and the lower the overall normalization.

 This third ladder rung corresponds to a picture wherein either the emitting gas is evenly distributed everywhere within the emission halo and is volume-filling, or the emitting gas is confined to small clouds but the clouds themselves are numerous and evenly distributed throughout the volume, like a mist. \citet{Mohapatra2022a} found that an emission-weighted projection of a second-order VSF tends to steepen the slope, in addition to reducing the overall normalization, \emph{unless} the gas is clumpy rather than volume-filling, which instead flattens the slope in projection because turbulent eddies along the LOS do not fully cancel with each other if only a handful of clumps are pierced by the LOS. These are exactly the two pictures we isolate with rungs 2 and 3 of the projection ladder, and we find the same results of making the VSF slope more shallow in rung 2, which represents a clumpy gas distribution, and steepening the slope in rung 3, which represents a more volume-filling distribution. \citet{Fournier2025} found similar results for the cold (clumpy) and hot (volume-filling) VSFs in a galaxy cluster core simulation with AGN feedback, and \citet{Xu2020} derives the slope of the projected VSF for both thin and thick slabs of turbulent gas, finding the projected VSF has a steeper slope when the slab of gas is thick, i.e., volume-filling.

Finally, the dark red solid curve in Figure~\ref{fig:projection_ladder} shows the VSF calculated from the density-squared-weighted average LOS velocity and projected separations, the final step we perform that mimics observations of VSFs from emission lines. The slope below the injection scale has shifted back toward $2/3$, indicating that perhaps the distribution of the gas is somewhere between a single cloud with unknown depth that flattens the slope (rung 2) and a volume-filling phase that steepens the slope (rung 3). We have not cut on gas temperature, so both cold clumpy material and warmer volume-filling phases are being mixed into the velocity map from which we calculate the VSF. This means that we may have both a steepening effect and a flattening effect in the projected VSF that end up canceling to produce a similar slope as the 3D VSF (see \S\ref{subsec:2D_VSF_temp} for VSFs calculated from different gas phases separately). The reduction in normalization compared to the full 3D VSF is due to a combination of the factor of 1/3 from reducing the velocity dimensionality (rung 1) and the reduction in power from averaging the LOS velocities (rung 3). By weighting the average LOS velocity by the density-squared, which peaks in the region immediately surrounding the galaxy, we are effectively sampling a smaller depth along the LOS than the size of the full box, so we do not lose as much power in the VSF as in rung 3. Appendix~\ref{appdx:projection_ladder} further discusses the dependence of the projected VSF on the depth along the line of sight.

\begin{figure}
  \centering
  \includegraphics[width=\linewidth]{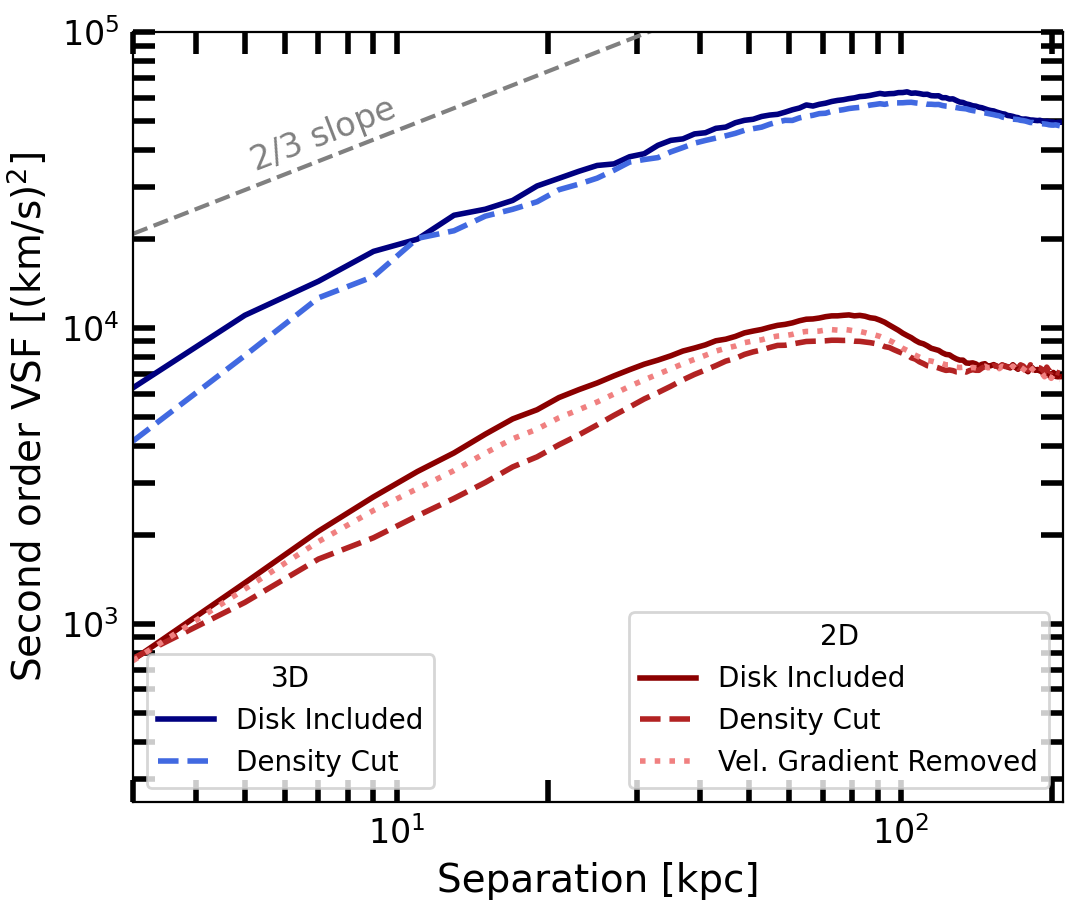}
  \caption{Second order VSFs in 3D (blue) and 2D (red). The solid lines indicate the VSF measured from all gas in the simulation, including the central galaxy disk. The dashed lines indicate the resultant VSFs after removing the disk using a density cut. The light red dotted line in 2D represents the disk removed by subtracting a velocity gradient from the data (see Figure~\ref{fig:velgrad}). The gray dashed line represents a Kolmogorov $2/3$ slope. The projection and calculation effects introduced when producing the 2D VSFs result in a decrease in both  normalization and turnover location when compared to the 3D VSF.}
  \label{fig:VSF_disk_removal}
\end{figure}

For the remainder of the paper, we focus on the full 3D VSF and the density-squared-weighted projected VSF, at the two ends of the ladder. Figure~\ref{fig:VSF_disk_removal} compares the second-order VSFs calculated in 3D (blue) and calculated from the 2D line-of-sight density-squared-weighted velocity projections (red). In both 2D and 3D, we show the VSF calculated with (solid) and without (dashed or dotted) the galaxy disk, using either the density cut removal method (dashed) or fitting and subtracting a velocity gradient from the disk in the 2D projected velocity map (dotted). In all cases, we use a box size of 200 kpc and a resolution of 2.2 kpc, for consistency of comparison. The 3D VSFs are calculated with a random selection of pairs of points while the 2D VSFs are calculated using all pairs of pixels in the projected image. Removing the disk decreases the normalization of the VSFs in both 3D and 2D, and in 2D shifts the location of the turnover to slightly smaller scales, but in both cases this is not a strong effect. The method of disk removal in 2D also does not appear to have a strong effect on the VSF. This is a little surprising, since \citet{Chen2023b} find that removing a velocity gradient slightly flattens the VSF, especially at large scales near the driving (turnover) scale and can thus shift the precise location of the turnover. We do not find such a strong effect on the VSF here, possibly because the disk occupies only a small fraction of the total image in Figure~\ref{fig:VSF_disk_removal} from which the VSF is calculated.

The above ladder stepping from 3D to emission-mimicked 2D VSFs explains the normalization and slope differences between the 3D and 2D VSFs, but the 2D VSFs also show a shift in the location of the turnover that is not explained purely by projection (e.g., in Appendix~\ref{appdx:projection_ladder}, the turnover location for the idealized velocity fields is always in the same location for every rung of the ladder). The average velocity in the 2D VSF is weighted by density-squared to mimic emission, and the gas density is higher in the center of the halo near the galaxy disk (see Figure~\ref{fig:diskprojs}) so the velocities near the center of the halo dominate in the average. We saw in Figure~\ref{fig:radial_bins} that the turnover of the VSF is located at smaller separations in the inner regions of the halo, so weighting by density, and thus preferentially the inner CGM, is likely causing the turnover location to be smaller in 2D than in 3D. These results suggest that the driving and strength of turbulence inferred from observed projected velocity maps may not fully represent the true, underlying turbulence in 3D, and is instead biased toward the location and velocity of the gas that is contributing to the emission.

\subsection{Effects of Measurement Box Size and Resolution on 2D VSFs}\label{subsec:2D_VSF_res_box}

Here, we perform the same analysis as in 3D to determine the effect of the size of the measurement region (``box size") on the turnover location of the 2D projected VSFs. We measure the location of the VSF's turnover or flattening point by finding the location of the maximum value of the VSF, and do this for multiple differently sized boxes, all centered on the center of the galaxy.

\begin{figure}
  \centering
  \includegraphics[width=\linewidth]{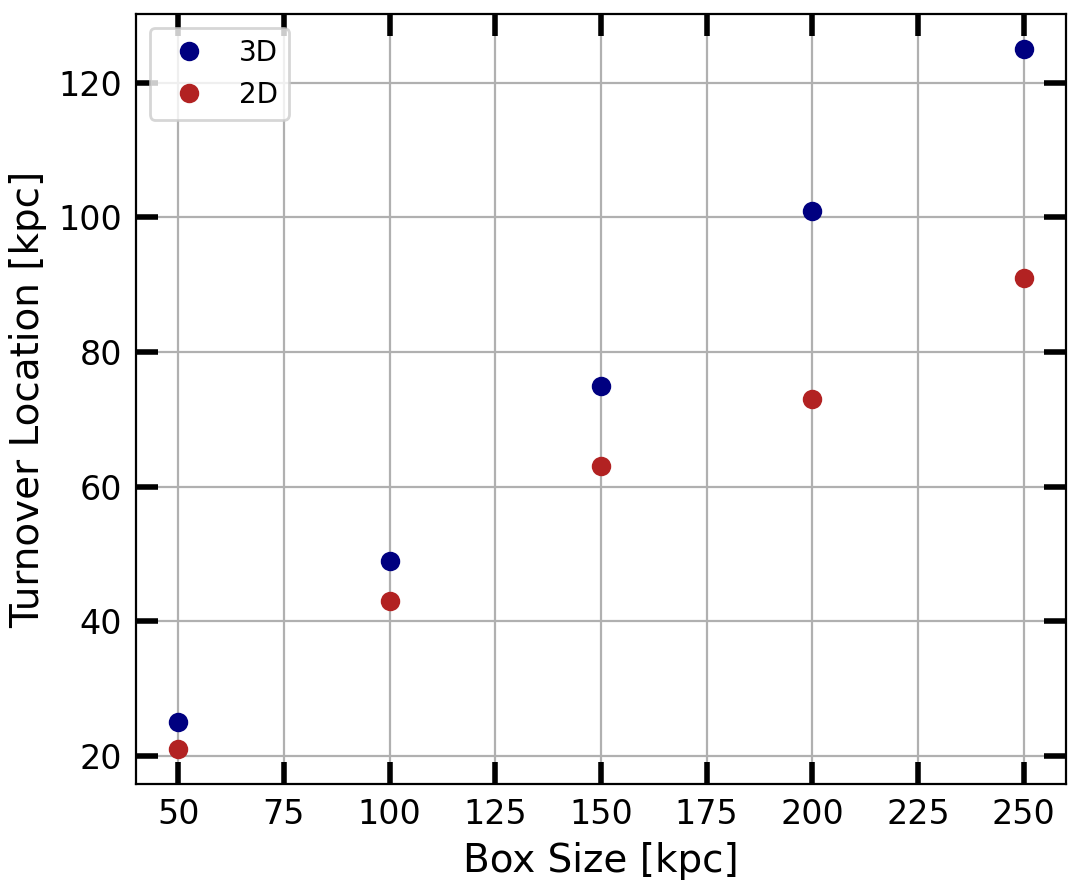}
  \caption{The blue points represent the location where the turnover of the 3D second order VSF plots occurs as a function of box size, demonstrating the effect that increasing box size has on the inferred driving scale of the turbulence. The red points illustrate the same effect for the 2D projected VSFs to compare these two, demonstrating that the 2D and 3D second order VSFs are similarly affected by box size.}
  \label{fig:box_peakloc}
\end{figure}

Figure~\ref{fig:box_peakloc} shows the turnover location of both the 2D and 3D VSFs as a function of increasing box size. We find that box size affects the 2D VSF in the same manner as in 3D: the location of the VSF's turnover increases to larger values as the size of the box increases. As before with 3D, this means that inferring the driving scale of turbulence from the flattening or turnover location of the 2D VSF needs to be done carefully, as it may be more closely related to the size of the region used to calculate the VSF (i.e., the size of the emission halo). For larger boxes, the turnover location of the 2D VSFs deviates from the linear slope seen in 3D at all box sizes. This may be because the 2D VSFs are density-squared-weighted and the gas density is highest at the galaxy in the center of the box, so the outskirts of the larger boxes do not affect the 2D VSF as strongly as they do in 3D, which is not weighted by gas density.

Next, we explore the impact of resolution on the 2D VSFs as we did in 3D (see Figure~\ref{fig:3D_resolution}). We test two different methods of varying the resolution: using progressively larger pixels like in 3D, and using a Gaussian smoothing function with different standard deviation $\sigma$ values applied to the highest-resolution pixel map. The former method gives a direct comparison to the resolution effects in 3D, and the latter mimics the effects of observational seeing with a large point-spread function (PSF) by blending small-small structures together.

\begin{figure*}
  \centering
  \includegraphics[width=\linewidth]{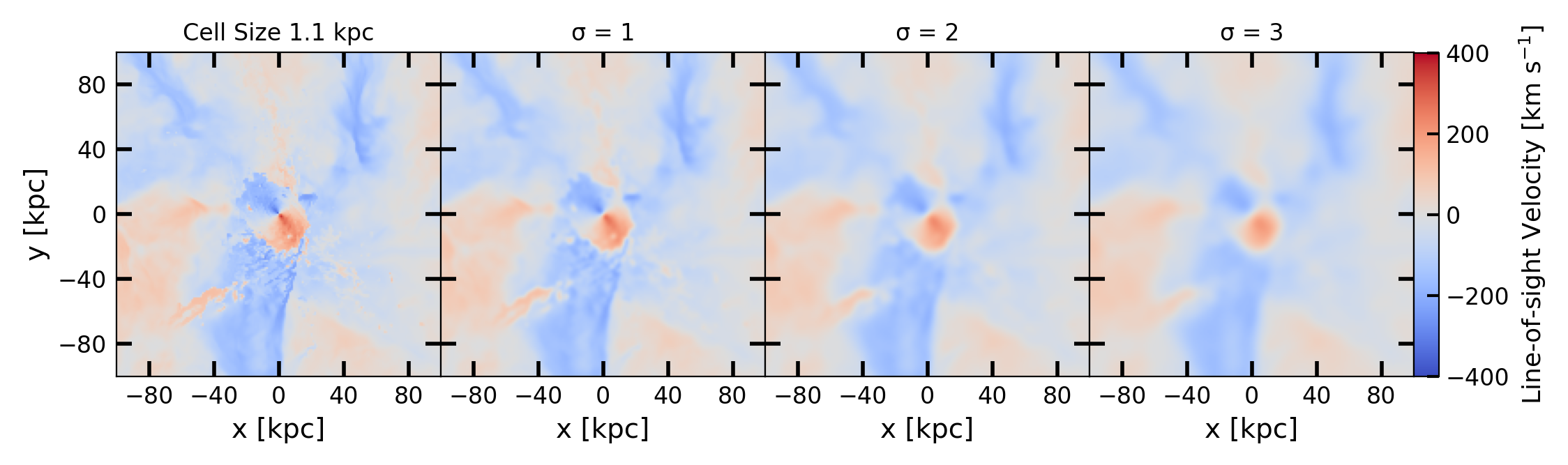}
  \caption{\emph{Left panel:} Density-squared-weighted mean LOS velocity of all gas of the simulated galaxy at maximum resolution. \emph{Right panels:} Subsequent panels represent increasing degrees of smoothing by increasing the standard deviation $\sigma$ (represented in number of pixels) within a Gaussian smoothing kernel. Broader smoothing removes more of the small-scale structure in the image.}
  \label{fig:gaussian_blur}
\end{figure*}

Figure~\ref{fig:gaussian_blur} demonstrates the effect of three different $\sigma$ smoothing values on the velocity map. The leftmost panel shows the density-squared-weighted average LOS velocity map at the highest resolution, and the panels moving to the right show progressively larger $\sigma$ values used in the Gaussian smoothing applied to the map in the left panel. The units of $\sigma$ are here given in number of image pixels, so that the Gaussian kernel with $\sigma=1$ encapsulates groups of roughly 4 pixels, $\sigma=2$ groups regions of roughly 16 pixels, and $\sigma=3$ groups regions of roughly 27 pixels. In this way, the $\sigma$ values can be matched up with the different image resolutions: cell sizes of $2.2$ kpc means each pixel represents the average of 4 of the highest-resolution $1.1$ kpc cells, which is similar to the $\sigma=1$ smoothing. Cell sizes of $4.4$ kpc contain 16 of the highest-resolution cells, similar to the $\sigma=2$ smoothing, and so on. The biggest difference between the two methods of exploring resolution is that in the first method, we extract the data from the simulation on the lower-resolution grid (which averages combined cells) \emph{before} creating the projection, so the resolution is lower along the line of sight as well as in the resulting 2D velocity map image, mimicking different simulation resolutions. The Gaussian smoothing method is applied \emph{after} creating the 2D velocity map from the fiducial resolution grid, mimicking the effects of observational seeing blurring out small details.

\begin{figure}
  \centering
  \includegraphics[width=\linewidth]{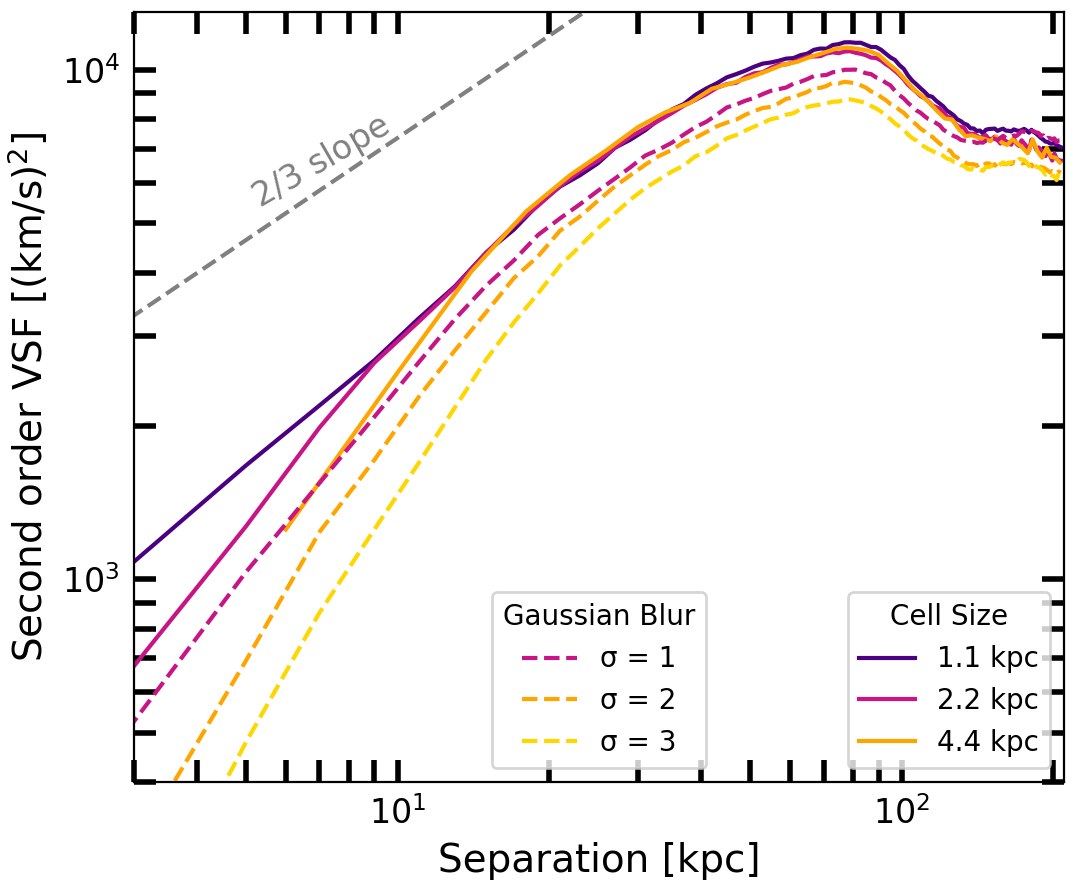}
  \caption{The second-order VSF calculated from density-squared-weighted average LOS velocity maps in 2D, using different simulation resolutions (solid curves) and by applying different levels of Gaussian smoothing to the map (dashed curves). The darkest purple line represents the highest resolution at a cell side length 1.1 kpc, and subsequent lighter colors represent increasing cell size and thus decreasing resolution, with the orange line at the worst resolution of cell side length 4.4 kpc. The pink dashed line represents Gaussian smoothing with $\sigma=1$, with lighter colors showing higher levels of smoothing up to $\sigma=3$. The matching colors between the solid and dashed curves represent comparable levels of resolution or Gaussian smoothing. Increasing Gaussian smoothing decreases normalization more significantly than changing the simulation resolution.}
  \label{fig:VSF_sigma}
\end{figure}

Figure~\ref{fig:VSF_sigma} shows the second-order VSFs measured from the 2D projected velocity images with different grid resolution levels (solid curves) and different Gaussian smoothing levels (dashed curves). Changing the grid resolution does not strongly affect the slope, normalization, or location of the turnover in the VSF, but higher degrees of Gaussian smoothing reduce the normalization of the VSF and steepen the slope on scales below approximately $10\times$ the size of the smoothing kernel. \citet{Chen2023b} and \citet{Chen2024} found similar effects of smoothing: they report a steepening of the VSF on scales smaller than $\sim10-20$ times the full width of the instrument PSF. This suggests that our Gaussian smoothing process mimics the effects of observational seeing well.

\subsection{VSFs at Different Gas Temperatures}\label{subsec:2D_VSF_temp} 

Emission observations obtain velocity maps for specific emission lines, such as [\ion{O}{2}] or [\ion{O}{3}], which typically trace gas of a specific temperature range where the ionization fraction of these ions is large. Because of this, emission lines do not necessarily trace the turbulent structure of all the gas, but rather of only a specific phase. To more closely match typical observational velocity maps, we bin the gas in the simulated CGM by temperature to ranges that typically represent low-ionization state gas ($T<10^{4.5}$ K), mid-ionization state gas ($10^{4.3}$ K $<T< 10^{5.5}$ K), and high-ionization state gas ($T>10^{5.5}$ K). The majority of observations that are currently feasible with ground-based IFUs trace the low-ionization-state gas that emits at optical wavelengths, while the mid-ions and high-ions may be observable in emission with upcoming UV space telescopes \citep{Saeedzadeh2025}, such as Aspera \citep{Chung2021,Lochhaas2026}.

\begin{figure*}
  \centering
  \includegraphics[width=\linewidth]{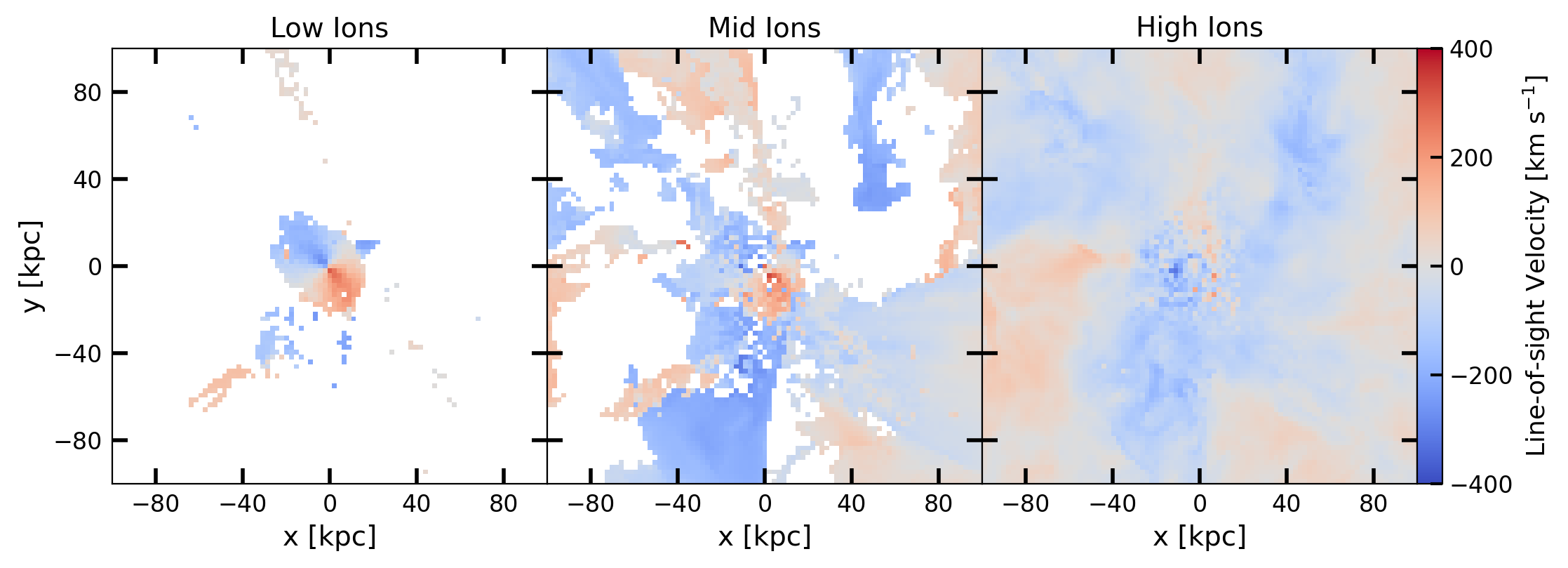}
  \caption{Density-squared-weighted mean LOS velocity of the simulated galaxy and CGM cut at different temperature ranges corresponding to typical ionization states. \emph{Left:} Gas with $T < {10}^{4.5}$ K, corresponding to typical low ions like [\ion{O}{2}] and [\ion{O}{3}]. \emph{Middle:} Gas with ${10}^{4.3}$ K $< T < {10}^{5.5}$ K, corresponding to typical mid-ions like \ion{C}{4} and \ion{Si}{4}. \emph{Right:} Gas with $T > {10}^{5.5}$ K, corresponding to typical high ions like \ion{O}{6} and \ion{N}{5}.}
  \label{fig:temp_cut_vels}
\end{figure*}

Figure~\ref{fig:temp_cut_vels} shows the disk and CGM gas in different temperature bins corresponding to low, mid, and high ions. Because the disk is mostly made up of cool and cold gas, we remove the rotation velocity of the disk using the velocity gradient fit method described in Section~\ref{subsec:projections} when calculating the low-ion VSF. For the mid and high ions, the disk barely contributes to the velocity map and so we do not remove its velocities before calculating the VSFs. Note that the coldest gas occupies a significantly different spatial location than the warm and hot gas: the cold gas is primarily confined to the disk of the galaxy, while the warm and hot gas are primarily outside of the disk. The spatial separation, as well as the fact that disk kinematics dominate in the cold phase, indicate that the cold and warmer gas phases are significantly de-coupled. Therefore, we do not necessarily expect the gas phases to have similar VSFs or turbulent properties like they perhaps might in small-scale multiphase turbulence simulations \citep[e.g.,][]{Mohapatra2022a}.

\begin{figure}
  \centering
  \includegraphics[width=\linewidth]{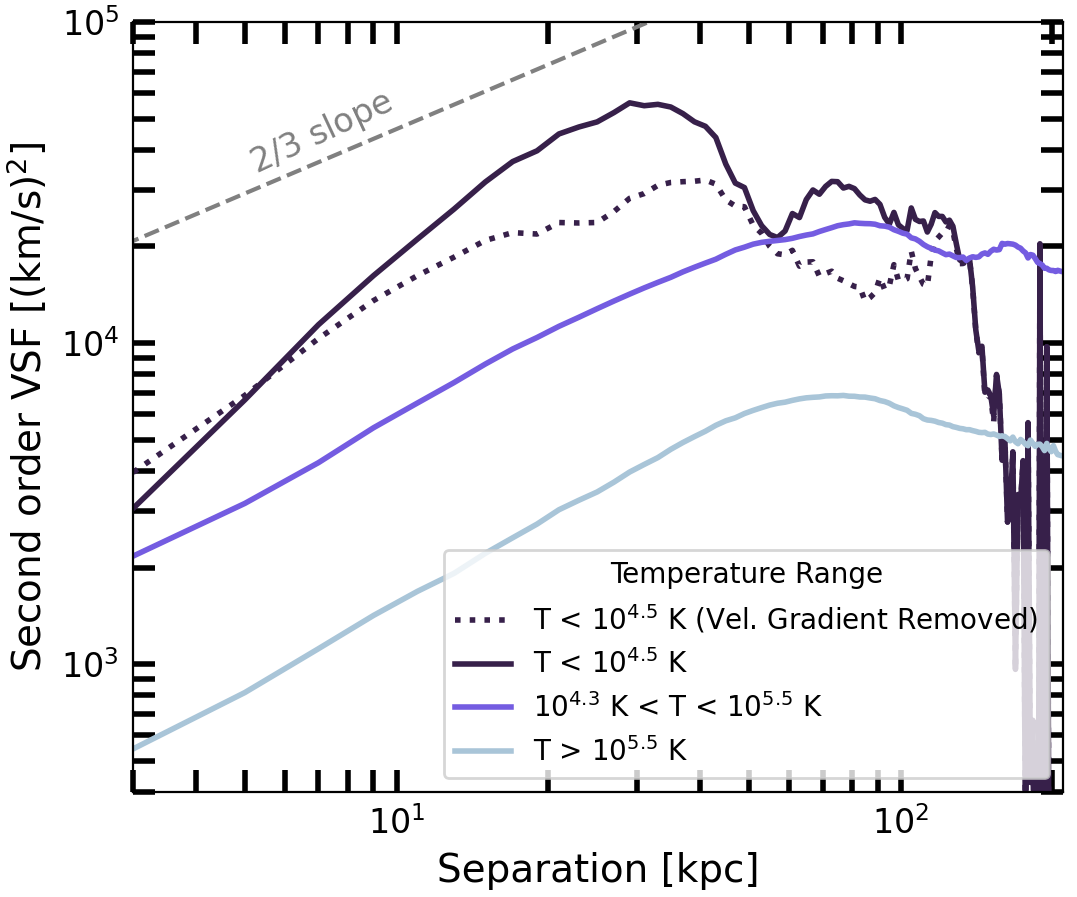}
  \caption{Second order velocity structure functions of 2D projections following the temperature cuts described in figure \ref{fig:temp_cut_vels}. The dark blue line represents the low ion temperature range in which $T < {10}^{4.5} K$. Since the low ion region is dominated mainly by the galaxy disk, the dark blue dotted line represents the same temperature range after removing the disk using a velocity gradient fit described in Section \ref{subsec:projections}. The purple line represents the VSF created using a temperature range of ${10}^{4.3} K < T < {10}^{5.5} K$, and the light blue line uses a range of $T > {10}^{5.5} K$. The 2/3 Kolmogorov scaling law slope is represented in the gray dashed line as a reference point.}
  \label{fig:temp_cut_VSFs}
\end{figure}

Figure~\ref{fig:temp_cut_VSFs} shows the second order VSFs calculated from the velocity maps in the different temperature bins. We show the lowest temperature VSFs both with (solid) and without (dotted) the disk. The mid and high ion VSFs are consistent with the expected $2/3$ Kolmogorov slope, however the low ion VSF including the disk is much steeper. This is due to the strong contribution of disk rotation to the velocity field, as when the disk is removed by subtracting the velocity gradient, the slope flattens to closer to $2/3$. This is consistent with the findings of \citet{Chen2023b} that neglecting to remove a large-scale velocity gradient produces a steeper slope in the VSF. However, the effect that we see here is much stronger than in \citet{Chen2023b}, likely because the low-ion temperature selection restricts the velocity image to essentially only the disk, with strong rotation. In \citet{Chen2023b}, the emission halos are more widespread than just the disk of the galaxy, and the velocity gradient is not as dominant as we find here.

The normalization and turnover locations of the VSFs in different temperature ranges are quite different. The coldest gas, which is almost entirely disk gas, has the highest normalization and smallest turnover location, which can be explained by fast velocities and the small scales of the disk. The highest-temperature gas has the lowest normalization, with a turnover location between the mid- and low-temperature gases. The highest-temperature gas in the simulation is primarily volume-filling material near the virial temperature \citep{Lochhaas2021}, which is denser closer to the galaxy. This could explain the intermediate driving scale: since the velocity maps are density-weighted, the VSF is weighted more towards smaller scales where the density is higher and the driving occurs on smaller scales, as we found in Figure~\ref{fig:radial_bins}. The low overall normalization of the VSF in the highest-temperature gas is likely due to power lost from averaging along the LOS, since this phase is most volume-filling (see \S\ref{subsec:proj_effects}). The mid-temperature gas in the FOGGIE simulations is a mixture of the extended gas disk \citep{Trapp2025} and inflowing streams (Lochhaas et al., in prep.), which occupy a larger area of the velocity map than just the disk and have turbulence likely driven by the process of accretion itself, which operates on large scales and could be responsible for the largest turnover location in this phase of gas \citep{Goldner2025}. The clumpier nature of the mid-temperature gas means less power is lost in the VSF to LOS averaging, so its normalization is higher than the high-temperature gas.

The [\ion{O}{2}] and [\ion{O}{3}] emission lines used to measure CGM VSFs in \citet{Chen2023b} and \citet{Chen2024} are generally expected to be tracing gas in our ``low ion" temperature range, $T\sim10^{4-4.5}$ K. Figure~\ref{fig:temp_cut_vels} shows that this temperature range selects nearly exclusively the galaxy disk, with little CGM material. However, the extents of the [\ion{O}{2}] and [\ion{O}{3}] observed halos are 30--40 kpc in radius from the central quasar, suggesting that emission from these ions is not confined to the galaxy disks, which are likely smaller. This points to a limitation of the FOGGIE simulations at $z\sim0$, and especially for Maelstrom: the CGM is not populated with very much cold material \citep{Trapp2025}. In addition, radiation from the quasars in the observed host galaxies are likely responsible for photoionizing the cool gas and producing the [\ion{O}{2}] and [\ion{O}{3}] emission, and the FOGGIE simulations do not include quasar feedback or any local sources of radiation, so this is not a direct comparison.

\section{Discussion} \label{sec:discussion}

\subsection{Caveats} \label{subsec:caveats}

In this work, we have focused on only a single simulation snapshot of a single simulated galaxy. Because we are most interested in the comparison between 3D and emission-mimicked 2D VSFs, we do not explore the properties of the turbulence itself or how it depends on redshift or galaxy properties. As a quick check, we examine the emission-mimicked 2D projected VSFs along the two other orthogonal projection directions for this $z=0$ snapshot of Maelstrom, and find that the VSFs among them are very similar, within 5\%. This suggests there may not be any structures with a preferred orientation in the CGM of Maelstrom, unlike the findings of \citet{Fournier2025}, where there are significant differences in the VSFs across different projection orientations and at different time snapshots in simulations of cool gas filaments located in a galaxy cluster core with AGN feedback.

We do not explore short-time variations in the turbulence, but we examine the $z=1$ snapshot of the same galaxy, Maelstrom (not shown). There, we do find differences in the 2D VSFs calculated from different projection directions, up to a factor of a few on the largest scales. At $z=1$ in the FOGGIE simulations, the CGM is more dominated by anisotropic bulk inflows and outflows than at $z=0$ (Lochhaas et al., in prep.), which can have a significant effect on the VSF depending on viewing angle. In addition, there is far more cold, clumpy gas outside of the disk in the CGM at $z=1$ in FOGGIE \citep{Augustin2025}, which is generally less symmetrically-distributed than the volume-filling phase. These effects together can make the projected VSF highly dependent on viewing angle at $z=1$.

We have not attempted to remove any bulk gas motions of inflows, outflows, or rotation outside of the disk. If these motions contribute strongly to the CGM velocity, they likely impact the VSFs as well. Such large-scale bulk flows would appear in the larger scales of the VSF, and so could affect the large-scale slope or the shape and location of the turnover. They could also contribute to making the turbulence anisotropic, which we also do not explore. If inflow regions have different turbulent properties than outflow regions, then mixing them together in a single VSF may also be mixing VSFs with different slopes, normalizations, and/or turnovers. Calculating VSFs for regions of gas with different radial velocities to separate radial flows is beyond the scope of this work, but we caution that the reported VSFs are likely representing a mix of processes and flows in the CGM, dominated by whichever has the strongest turbulence and the largest driving scale \citep{Yoo2014}.

\subsection{Comparison to Studies of Observed Turbulence} \label{subsec:obs_compare}

Velocity structure function measurements of turbulence beyond the ISM are limited to a handful of cases where gas emission is strong and widespread enough to obtain velocity information across a variety of scales. This includes H$\alpha$ filaments in galaxy clusters \citep{Li2020,Ganguly2023}, ram-pressure-stripped tails of jellyfish galaxies \citep{Li2023,Ignesti2024}, and, most relevant for comparison with our present work, the emission halos of quasar-host galaxies \citep{Chen2023b,Chen2024,Chen2025}. The FOGGIE simulated galaxy we explore here does not include an implementation for quasar feedback, and we examine the $z=0$ simulation output rather than $z\sim0.5-1$ like these observed systems, so we do not perform a direct, apples-to-apples comparison. However, the halo mass of FOGGIE's Maelstrom galaxy at $z=0$ is $10^{12}M_\odot$, similar to the expected halo masses of the quasar hosts, and we focus on turbulence in the inner CGM, similar to the location probed by the observed emission halos. Rather than compare the properties of turbulence across disparate systems, we focus here on the relationship between the 3D ``true" VSF and a mock-observed (projected and density-squared-weighted) VSF.

In general, we find that projection does not have much impact on the slope of the VSF, but can shift the location of the turnover or flattening point that is often used as an indicator of the turbulent driving scale. The location of the turnover can be related to the size of the region used to measure it (see Appendix~\ref{appdx:box_size}), but we found in Figure~\ref{fig:VSF_disk_removal} that the 2D VSF turned over at a smaller scale even when the same size of measurement region was used in both 2D and 3D. This can be attributed to the density-squared weighting of emission that prioritizes the inner CGM where densities are higher and the VSF turnover is at smaller scales (Figure~\ref{fig:radial_bins}). The combination of this and other projection effects (see \S\ref{subsec:proj_effects} and Appendix~\ref{appdx:projection_ladder}) makes it difficult to determine how the turnover of the projected VSF relates to the ``true" turbulence driving scale, so we stress that caution must be taken when inferring turbulent driving and sources from observed VSFs. Indeed, \citet{Chen2023b} find VSFs that flatten at scales around $\sim10-30$ kpc for emission halos of diameter $\sim60-70$ kpc, suggesting that the larger turnover scales close to half of the emission diameter may be affected by the size of the measurement region (see Appendix~\ref{appdx:box_size}).

\citet{Chen2023b} and \citet{Chen2024} found that noticeable velocity gradients across the emission halo can lead to steeper slopes of the measured VSF, and they fit and remove gradients. The most obvious velocity gradient we find in the Maelstrom galaxy at $z=0$ is due to the disk rotation, which we remove either by removing the disk gas from the calculation entirely or by fitting and removing the velocity gradient in the disk region (\S\ref{subsec:proj_effects}). However, we do not find this makes a substantial difference to the VSF when all gas is used to measure it (Figure~\ref{fig:VSF_disk_removal}), because the strong velocity gradient exists only in the relatively small disk while the broader CGM is more well-mixed. When we restrict the measurement of the VSF to only cool gas that would conceivably be producing [\ion{O}{2}] or [\ion{O}{3}] emission, the measurement region is restricted essentially entirely to the disk, and then we do see a strongly steeper slope unless the velocity gradient is fit and removed (Figs.~\ref{fig:temp_cut_vels} and~\ref{fig:temp_cut_VSFs}).

As an additional test of the impact of velocity gradients, we also examine the $z=1$ output of Maelstrom (not shown), which exhibits a velocity gradient not just in the galaxy disk, but in its CGM as well. At higher redshifts $z\gtrsim1$ in FOGGIE, the galaxies exhibit more filamentary and less-spherically-symmetric accretion that generates CGM velocity fields with more bulk movement, leading to broad velocity gradients across $\sim100$ kpc. We find similar results: fitting and subtracting off the velocity gradient flattens the VSF and brings it in line with a Kolmogorov $2/3$ slope.

\subsection{Comparison to Other Studies of Simulated Turbulence} \label{subsec:sim_compare}

Using small-box simulations of driven turbulence that represent small patches of multiphase gas in the CGM or ICM, \citet{Mohapatra2022b} and \citet{Mohapatra2022a} determine how the measured VSFs depend on the phase of the gas observed, the type of turbulent driving, and projection effects. They find that, in the case where the turbulence is primarily solenoidal and magnetic fields are not included, the hot and cold gas phases have similar VSFs. We find a similar result for the mid- and high-temperature gas (Figure~\ref{fig:temp_cut_VSFs}), except that we find a higher overall normalization of the VSF for the mid-temperature gas than the high-temperature gas. The multiphase geometry of gas in our cosmological simulated halo is likely quite different than that represented by a small-box simulation, so it is not surprising there are some differences. In particular, nearly all of the coldest gas in Maelstrom is confined to the disk of the galaxy, so we do not compare its VSF to the cold gas in the small boxes of these studies. \citet{Wang2021} also find that the cold and hot gas phases in a simulation of a cluster core with AGN feedback can exhibit different slopes, but that magnetic fields can enhance coupling between phases.

\citet{Fournier2025} also perform the experiment of Gaussian smoothing on the velocity field to mimic the effects of observational seeing, as we do in Figs.~\ref{fig:gaussian_blur}-\ref{fig:VSF_sigma}. We find similar results to \citet{Fournier2025}: the VSF steepens near and below the scale that corresponds to the size of the smoothing kernel due to small-scale velocities being smeared together within the kernel.

\section{Conclusions} \label{sec:summary}

In this work, we use the FOGGIE cosmological zoom-in simulations with enhanced CGM resolution to determine how well emission-weighted projected VSFs capture CGM turbulent properties as compared to VSFs measured from full 3D information. We examined one Milky Way-like galaxy at $z=0$ from the FOGGIE suite and focused on the CGM by removing the galaxy disk through either a gas density cut (\S\ref{subsec:disk_removal}, Figure~\ref{fig:diskprojs}) or by fitting and subtracting off a gradient in the LOS velocity field (\S\ref{subsec:projections}, Figure~\ref{fig:velgrad}). We focus primarily on the location of the change in VSF slope, the ``turnover", which may be a tracer of the turbulence driving scale. Our main results are as follows:
\begin{itemize}
    \item Binning the simulation to larger cell sizes (worse grid resolution) has little impact on the slope or turnover location of the 3D VSF, but slightly reduces the overall normalization (Figure~\ref{fig:3D_resolution}).
    \item Measuring the 3D VSF in larger boxes has a strong impact on the normalization and turnover location (Figure~\ref{fig:3D_box_size}). This is partially because turbulence far from the galaxy has different properties, which is picked up by the larger box sizes (Figure~\ref{fig:radial_bins} and Appendix~\ref{appdx:many_driving}), and partially because using a too-small measurement box can generate a turnover in the VSF that is not located at the actual driving scale (Figure~\ref{fig:box_peakloc} and Appendix~\ref{appdx:box_size}).
    \item The 2D emission-weighted projected VSF has a lower normalization and smaller turnover location than the 3D VSF measured from the same box (Figure~\ref{fig:VSF_disk_removal}). The lower normalization is due to a combination of reducing the velocity information from 3D to 1D (line of sight velocity) and averaging velocities along the line of sight, and the smaller turnover location is due to the density-squared-weighting, which mimics emission, prioritizing the inner CGM where densities are higher and the turnover location is smaller.
    \item When the 2D projected velocity field is smoothed to mimic observational seeing (Figure~\ref{fig:gaussian_blur}), the VSF steepens on scales below $\sim10\times$ the size of the smoothing kernel and the overall normalization of the VSF decreases slightly (Figure~\ref{fig:VSF_sigma}).
    \item We split the 2D projected velocity field into gas temperature bins (Figure~\ref{fig:temp_cut_vels}) to mimic emission from particular lines that trace a specific phase of gas, and measure the VSF from each independently. The intermediate- and high-temperature gases have similar VSF shapes, with lower normalization in the high-temperature gas because it is more volume-filling, so more velocities are averaged together along the line of sight (Figure~\ref{fig:temp_cut_VSFs}). The coldest gas is primarily confined to the disk of the galaxy and shows a steep VSF unless the velocity gradient of the disk is removed, after which the slope becomes similar to the hotter gas phases.
\end{itemize}

Interestingly, the majority of our measured VSFs --- in 3D, in projection, at different resolutions or different gas phases --- have slopes close to the expected Kolmogorov slope of $2/3$ in the inertial range of scales. The simulated galaxy we examine at $z=0$ is fairly quiescent and its CGM gas is fairly symmetric, so the isotropic, subsonic and incompressible assumptions inherent to Kolmogorov turbulence may be satisfied.

However, our finding that the location of the VSF turnover may or may not actually correspond to the driving scale of the turbulence suggests extreme caution must be taken when interpreting observed VSFs. The measurement area for observed VSFs is limited by the extent of bright emission above the instrument's sensitivity, and if the turnover location is more closely related to the size of this measurement area than to the underlying turbulent properties, then it may not be appropriate to infer the source of turbulent driving from the VSF. Appendix~\ref{appdx:box_size} suggests a test using sub-sections of the data that could help determine if the turnover scale is ``real" or not.

Future observations with both optical and UV IFUs will be instrumental to further measurements of turbulence in the CGM gas. Obtaining both spatial and kinematic information in emission from the diffuse CGM is already possible in the optical with long exposure times, using MUSE or KCWI. The proposed next-generation NASA flagship, the Habitable Worlds Observatory, will have a UV IFU and significantly higher sensitivity limits than any existing UV telescope \citep{Burchett2025}. This will allow for deeper and more precise probes of turbulence and other CGM gas kinematics at lower redshifts that will require the kinds of simulation prediction carried out here to understand how to interpret.

\begin{acknowledgments}
The authors are grateful for the the FOGGIE collaboration, especially PI Molly Peeples and co-PIs Jason Tumlinson and Brian O'Shea, for providing the FOGGIE simulation data that was used in this work. Yuan Li and Brian O'Shea provided useful feedback and discussion that improved the quality of the text. Support for C.L. was provided by NASA through the NASA Hubble Fellowship grant \#HST-HF2-51538.001-A awarded by the Space Telescope Science Institute, which is operated by the Association of Universities for Research in Astronomy, Inc., for NASA, under contract NAS5-26555. J.S.B. acknowledges support from the Simons' Collaboration on Learning the Universe and the Leverhulme Trust. Support for H.B. was provided by the Northeastern University College of Science Research Co-op Fund. M.C.C. is supported by the Brinson Foundation through the Brinson Prize Fellowship Program.

Computations described in this work were performed using the publicly-available \textsc{Enzo} code (\href{http://enzo-project.org}{http://enzo-project.org}), which is the product of a collaborative effort of many independent scientists from numerous institutions around the world. Their commitment to open science has helped make this work possible.

Resources supporting this work were provided by the NASA High-End Computing (HEC) Program through the NASA Advanced Supercomputing (NAS) Division at Ames Research Center and were sponsored by NASA's Science Mission Directorate; we are grateful for the superb user-support provided by NAS.
\end{acknowledgments}

\software{\textsc{astropy} \citep{Astropy2013,Astropy2018},
          \textsc{matplotlib} \citep{matplotlib},
          \textsc{numpy} \citep{numpy},
          \textsc{scipy} \citep{scipy},
          \textsc{yt} \citep{yt}}

\begin{contribution}
H.B. led the analysis, developed all figures in the main body of the paper, and wrote parts of the methods and results sections. C.L. conceived the project idea and general analysis plan, co-mentored H.B. through the analysis, wrote the introduction, discussion, and parts of the methods and results sections and the comparisons to other work, and carried out the analysis for and wrote all appendices. J.S.B. co-mentored H.B. through the analysis and provided significant suggestions on both the analysis and paper text. M.C.C. provided significant discussion and suggestions that led to the creation of the ``projection ladder" in Section 4.2 and Appendix C.
\end{contribution}

\appendix
\section{Dependence of VSF Turnover on Box Size}
\label{appdx:box_size}

In this appendix, we use idealized velocity fields to determine the correlation of VSF turnover location with the size of the box used for measurement.

We begin by generating a cube of 3D turbulent velocity fields by sampling velocities from a specified power spectrum, following the method of \citet{ZuHone2016}. We start by defining a Kolmogorov spectrum for the power spectrum:
\begin{equation}
    P(k) = C_n k^{-11/3} e^{-(k/k_\mathrm{peak})^2}
\end{equation}
where $C_n$ is a normalization constant, $k_\mathrm{peak}$ is the wavenumber where the spectrum cuts off, representing the turbulence injection scale, and the power of $-11/3$ recovers the expected $-5/3$ power law scaling for Kolmogorov turbulence in the energy power spectrum, $E(k)=P(k)k^2$. We generate a wavenumber grid from $1/N_r$ to $N_r$, where $N_r=512$ is the resolution along one side of the cube, for each of $x$, $y$, and $z$ directions, then calculate the energy spectrum $E(k)$ everywhere on the grid. We set the normalization constant $C_n$ such that the 1D velocity dispersion of each of the $x$, $y$, and $z$ velocity fields will be $\sigma_v=50$ km/s:
\begin{equation}
    C_n = \frac{N_r^3\sigma_v^2}{\Sigma E(k)}
\end{equation}
where $\Sigma E(k)$ is the calculated energy power spectrum summed across the wavenumber grid. With $E(k)$ now fully specified, we generate a complex field for each of $x$, $y$, and $z$ on the grid with amplitude $\sqrt{2 E(k)}$ and a random direction:
\begin{equation}
    \Tilde{v_d} = \sqrt{2E(k)}e^{2\pi i R}
\end{equation}
where the subscript $d$ represents $x$, $y$, or $z$ directions and $R$ is a random number between 0 and 1. Finally, we take the Fourier transform of $\Tilde{v_d}$ with orthonormal normalization to obtain $v_d$.

\begin{figure*}
    \centering
    \includegraphics[width=\linewidth]{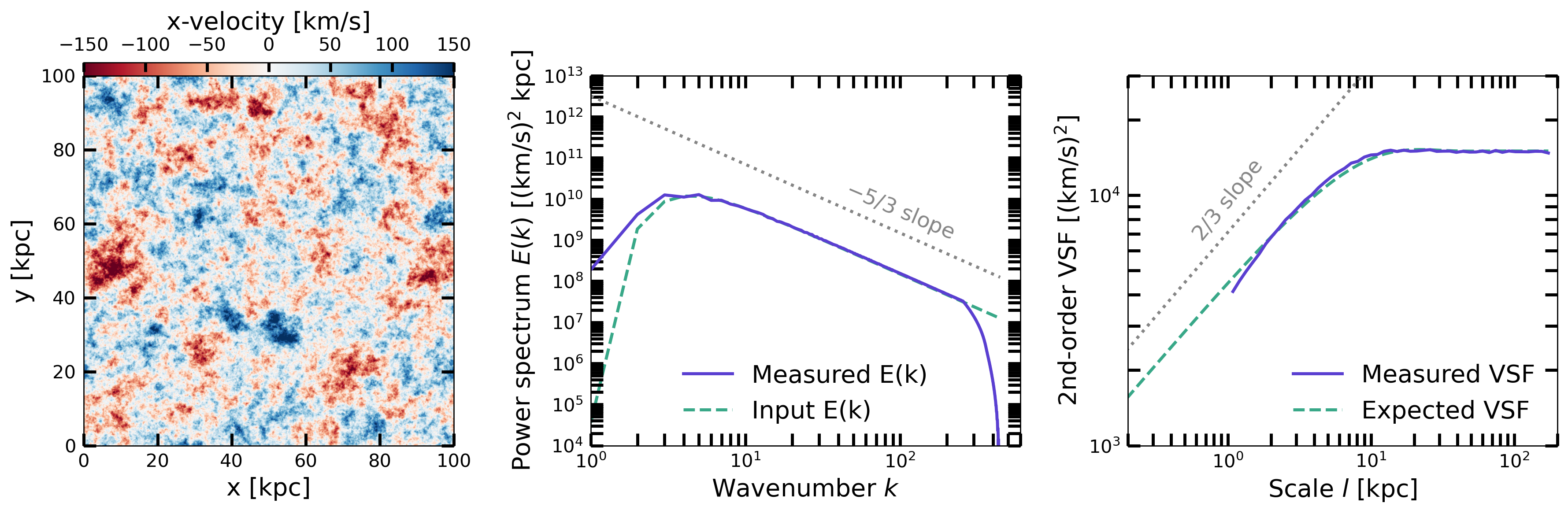}
    \caption{(\emph{Left:}) A slice of $v_x$ through the center of a cube generated with Kolmogorov turbulence with $k_\mathrm{peak}=4$ and $\sigma_v=50$ km/s. (\emph{Middle:}) 1D energy power spectra expected (teal dashed) and measured (solid purple) from this box. The measured spectrum matches the input spectrum well, with the expected $-5/3$ Kolmogorov slope. (\emph{Right:}) Second-order VSF expected (teal dashed) and measured (solid purple) from this box. The measured VSF also matches the expectation well, with a flattening scale of $\sim12.5$ kpc, or $1/8$ of the box, corresponding to $k_\mathrm{peak}=4$ as expected. The $2/3$ slope of Kolmogorov turbulence is evident.}
    \label{fig:ideal_box}
\end{figure*}

The result is a 3D cube containing random, isotropic Kolmogorov turbulence in each of the $x$, $y$, and $z$ velocities. We use a resolution of $N_r=512$ and set the 1D velocity dispersion to $50$ km/s, and assume the physical size this cube represents is $L=100$ kpc on a side. The left panel of Figure~\ref{fig:ideal_box} shows a slice of $v_x$ through the center of the generated cube. The middle panel shows the 1D energy power spectrum that is expected from this generation process, with a Kolmogorov slope of $-5/3$, as the dashed teal line. We measure the 1D energy power spectrum from the generated cube by first taking the Fourier transform (again with orthonormal normalization) of the velocity fields, then summing $(\Tilde{v_x}^2+\Tilde{v_y}^2+\Tilde{v_z}^2)/6$ in bins of $k$. The solid purple line shows this measured 1D energy power spectrum from the resulting box, which matches the expected spectrum well, ensuring that the velocity field generation process has worked as expected. There are deviations between the input and output spectra at large and small $k$, which represent finite volume and resolution, respectively, in the generated cube.

The expected second-order VSF is calculated from the input $E(k)$ as
\begin{equation}
    S_2(l) = C_n \int_0^\infty  \left(1 - \frac{\sin(\pi k l/2)}{\pi k l/2}\right) k^{-5/3} e^{-(k_\mathrm{peak}/k)^2} \mathrm{d}k
\end{equation}
where the normalization constant is given by
\begin{equation}
    C_n = \frac{6\sigma_v^2}{\int_0^\infty  k^{-5/3} e^{-(k_\mathrm{peak}/k)^2} \mathrm{d}k}
\end{equation}
In the above equations, $k$ is the wavenumber, $l$ is the physical scale on which the VSF is measured, and $S_2(l)$ indicates the second-order VSF.

We measure the second-order VSF from this idealized cube in a slightly different way than we do from the FOGGIE simulations. To ensure a good sampling of the inertial range, we use a method that ensures more pairs are selected at small separations than would be expected from truly random sampling. We start by choosing 50 log-spaced bins of pair separation $l$ between $5\times$ the resolution and $\sqrt{3}\times$ the cube side length (to account for the largest separation along the diagonal). We pick a random cell in the full cube, then generate a vector from this point with a random direction and length uniformly randomly chosen from the separation bin of interest, and use the cell the vector points to as the second point in the pair. We perform this pair selection procedure $10^4$ times for each separation bin and use these pairs to measure the VSF.

The right panel of Figure~\ref{fig:ideal_box} shows the expected and measured VSF from this generated cube. The two curves agree well with each other, and both show the $2/3$ slope expected for a second-order VSF under Kolmogorov turbulence. We expect the flattening point to occur at about half of $L/k_\mathrm{peak}$, where $L$ is the physical side length of the full cube \citep{ZuHone2016}. Indeed, the flattening point occurs around 12.5 kpc, which matches well for $L=100$ kpc and $k_\mathrm{peak}=4$. Having tested that the velocity field generation procedure works as we expect, with good correspondence between the expected and measured power spectra and second-order VSF, we now explore the effects of measuring the VSF from this same cube using differently-sized sub-regions.

We pick four boxes of different sizes as the sub-regions of the full cube. They have side lengths equal to $1/2,\ 1/4,\ 1/8,$ and $1/16$ the full cube side length. We randomly choose the location of each box within the cube, ensuring it does not cross over the edges, and calculate the second-order VSF from each box. We repeat this process 8 times for the boxes that are $1/2$ the cube side length and 64 times for the other box sizes, and then average the VSFs together for all boxes of the same size.

\begin{figure*}
    \centering
    \includegraphics[width=0.75\linewidth]{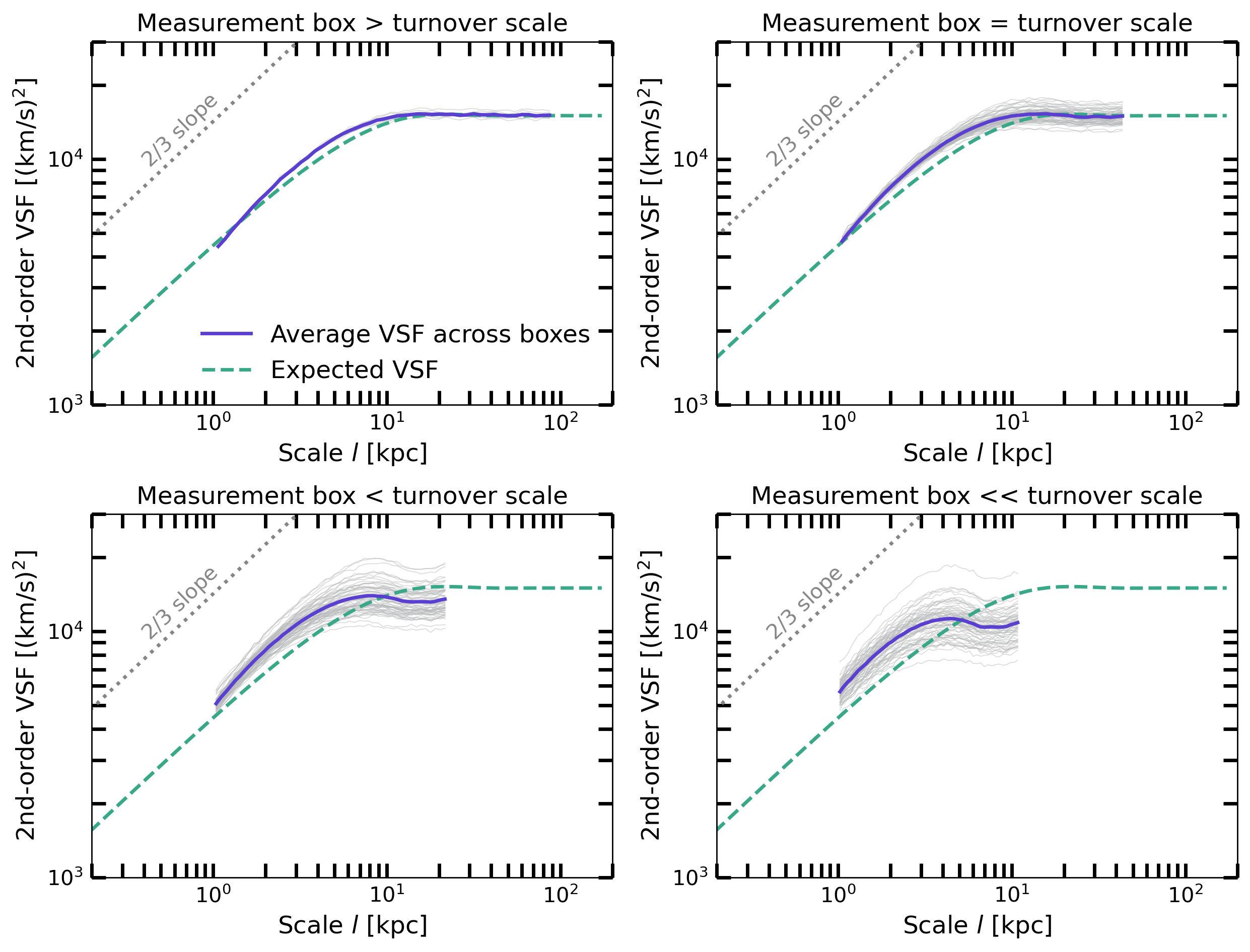}
    \caption{Each panel shows the second-order VSF measured from sub-regions of the full turbulent cube, decreasing in size of the sub-region from left to right and top to bottom. The faint gray curves show the VSFs from each of many randomly-selected sub-regions, while the solid purple curve shows the average among these. The dashed teal curve shows the expectation from the full cube. As the measurement region decreases in size below the scale of turbulence, the VSF shifts further away from the full cube's VSF and the flattening scale no longer accurately captures the scale of the turbulence injection.}
    \label{fig:VSFs_subboxes}
\end{figure*}

Figure~\ref{fig:VSFs_subboxes} shows the resulting VSFs from each sub-region box. The $1/2$-size box is shown in the top left, the $1/4$-size box in the top right, the $1/8$-size box in the bottom left, and the $1/16$-size box in the bottom right. Because the full cube contains turbulence generated with $k_\mathrm{peak}=4$, the $1/2$-size box is larger than this turnover, the $1/4$-size box is equal to the turnover scale, and the $1/8$ and $1/16$ boxes are smaller than the turnover scale. As the size of the box used to measure the VSF decreases, each individual box's VSF (faint gray curves) is noisier, but they all produce a flattening scale that shifts to smaller values, and for the smallest box, lower normalization.

To determine how the turnover or flattening scale of the measured VSF depends on the size of the sub-region used to measure it, we repeat the above process for several additional sub-region box sizes less than and greater than the turnover scale in the full cube. We calculate the VSF from $10^4$ random pairs per separation bin within each of 5--50 randomly selected sub-regions (we select fewer sub-regions for the larger sub-region box sizes to avoid too much potential for overlapping regions), and then average together the VSFs to produce one smooth VSF for each sub-region box size. We then find the scale at which this averaged VSF flattens by searching for the first scale where the log-log slope drops below a threshold of 0.08, which was determined empirically to match the VSF flattening scale well by-eye.

\begin{figure}
    \centering
    \includegraphics[width=0.5\linewidth]{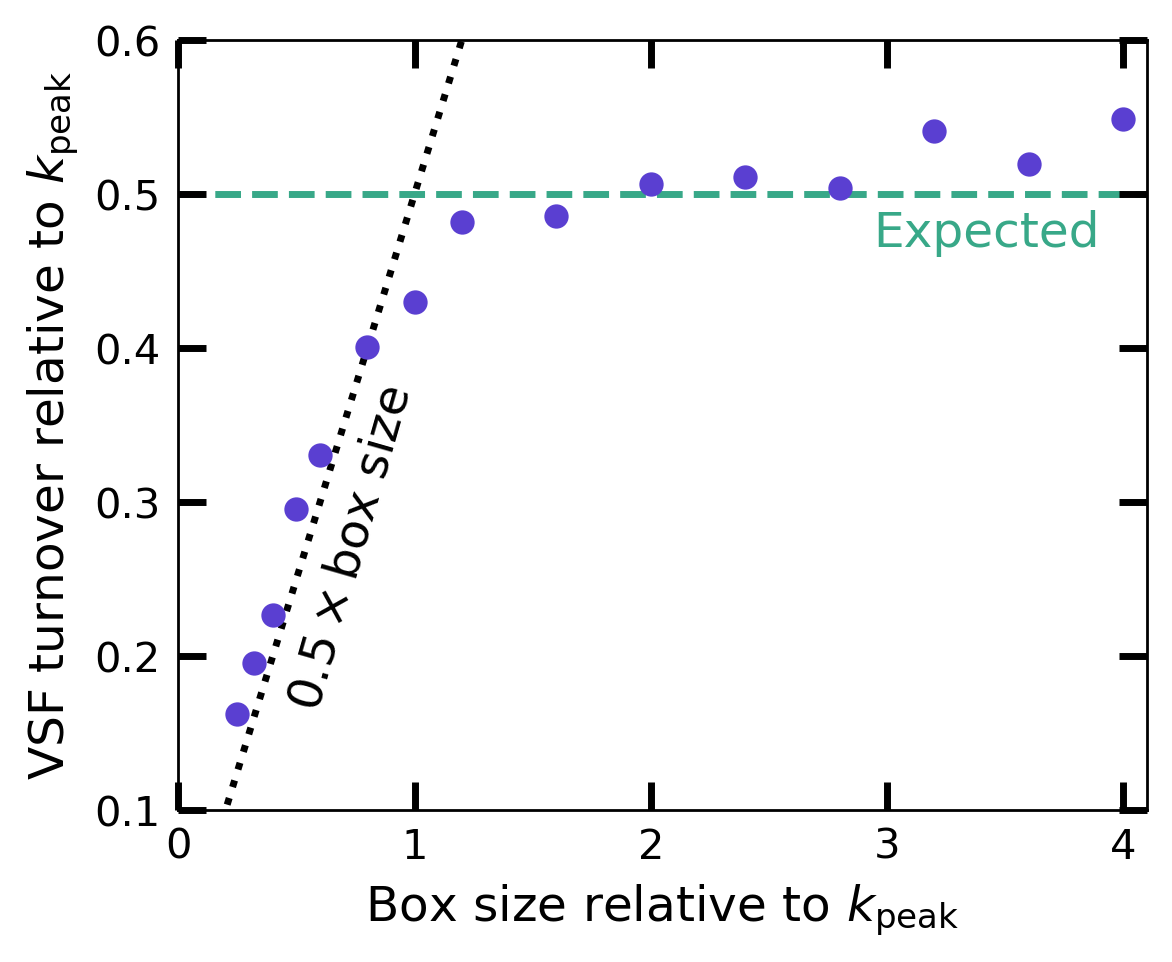}
    \caption{The location of the turnover, or flattening, scale of the second-order VSF measured from sub-region boxes of different sizes, as a function of the sub-region box side length. Both axes are scaled to $L/k_\mathrm{peak}$. Measurement regions larger than $L/k_\mathrm{peak}$ capture the ``true" turnover scale of the full cube, while measurement regions smaller than $L/k_\mathrm{peak}$ produce turnover scales of 1/2 the measurement region size.}
    \label{fig:ideal_boxsize_peakloc}
\end{figure}

Figure~\ref{fig:ideal_boxsize_peakloc} shows this measured VSF turnover location as a function of the box size of the sub-regions of the full cube from which the VSFs were measured. Both axes are scaled relative to $L/k_\mathrm{peak}$. In the full cube, the VSF turns over at $L/(2k_\mathrm{peak})$, which is marked as the horizontal teal dashed line. Sub-region box sizes larger than $L/k_\mathrm{peak}=1$ produce VSFs with turnover scales that match this expectation. However, if the box size is less than $L/k_\mathrm{peak}$, the turnover scale is less than the ``true" turnover scale, and is instead determined by the sub-region box size such that the flattening scale is $1/2$ the box side length.

Our results here are similar to those found by \citet{GarciaVazquez2023} in their Appendix A. While we perform this test using full 3D VSFs, they perform it in 2D with similarly-generated velocity fields as we do here, but additionally including a randomly generated fluctuating emissivity field from which they calculate the emissivity-weighted velocities. They find that measuring the VSF using smaller regions than $\sim10\times$ the velocity correlation length $r_0$ produces an apparent correlation length that is smaller than the ``true" correlation length in the full velocity field. We report our results in terms of box size $L$ relative to the driving wavenumber $k_{\rm peak}$, where the flattening of the VSF (when measured with a suitably large measurement region) occurs at a scale of $\frac{1}{2}L/k_{\rm peak}$. The correlation length $r_0$ of \citet{GarciaVazquez2023} appears to be roughly 1/5 of our flattening scale, and therefore 1/10 of the driving scale $L/k_{\rm peak}$. Thus, their finding that the measurement box size must be $\gtrsim10\times$ the correlation length to obtain an accurate measurement is equivalent to our finding that the measurement box size must be $\gtrsim1\times$ the driving scale to obtain an accurate measurement, and we are fully in agreement. We also find a wider dispersion in the VSF measured in smaller boxes, just as they do.

However, whereas \citet{GarciaVazquez2023} find the VSFs for the different measurement box sizes sit on top of each other (with missing information on larger scales in the smaller boxes), we find that the whole VSF shifts to the left in Figure~\ref{fig:VSFs_subboxes} as smaller measurement boxes are used. This disagreement may be due to the $\sigma^2$ normalization they perform while we do not, or the emissivity-weighting and 2D projection, which we also do not do here. Regardless, we agree with the general result that the turbulence injection scale, whether it is measured with the flattening of the VSF or the correlation length, cannot be accurately measured using too-small measurement regions.

In an astrophysical context, the results of this section indicate that the inferred ``driving scale" of turbulence, typically measured from the turnover or flattening point of the VSF, \emph{only accurately captures the true turbulent properties if the measurement region is larger than the driving scale}. If this is not the case, such as in an observation where emission can only be detected from a small region, then the turnover or flattening of the VSF \emph{does not contain any information about the driving scale of turbulence}, instead scaling only with the size of the measurement region. An observer could determine which regime their measurement falls into by taking smaller and smaller sub-sections of the data and determining if the turnover scale changes with the size of the region used to measure the VSF. If a constant value is approached with the larger regions, as in Figure~\ref{fig:ideal_boxsize_peakloc}, the observer could be confident that the turnover is accurately capturing the true driving scale of the turbulence.

This same principle also applies to the radial shells of different radii, as explored in Section~\ref{subsec:radial} and Figure~\ref{fig:radial_bins}. If the radius of a shell is smaller than the driving scale, then the flattening or turnover location of the VSF scales with the shell radius rather than containing information about the driving scale. We tested this to be the case by calculating the VSF in radial shells in the same idealized velocity fields (not shown) and found a similar trend as when using measurement boxes of different sizes, as in Figures~\ref{fig:VSFs_subboxes}-\ref{fig:ideal_boxsize_peakloc}. 

\section{Measuring VSFs from turbulence with many driving scales}
\label{appdx:many_driving}

In this section, we explore the impact of multiple driving scales affecting the measured VSF. We again use an idealized turbulent velocity field generated as described in Appendix~\ref{appdx:box_size}, but now we use different values of $k_\mathrm{peak}$ and different values of the velocity dispersion $\sigma_v$ within spherical shells. We also increase the box size to $L=200$ kpc, defined with $(x,y,z)=(0,0,0)$ in the center for ease of applying spherical symmetry.

\begin{figure*}
    \centering
    \includegraphics[width=\linewidth]{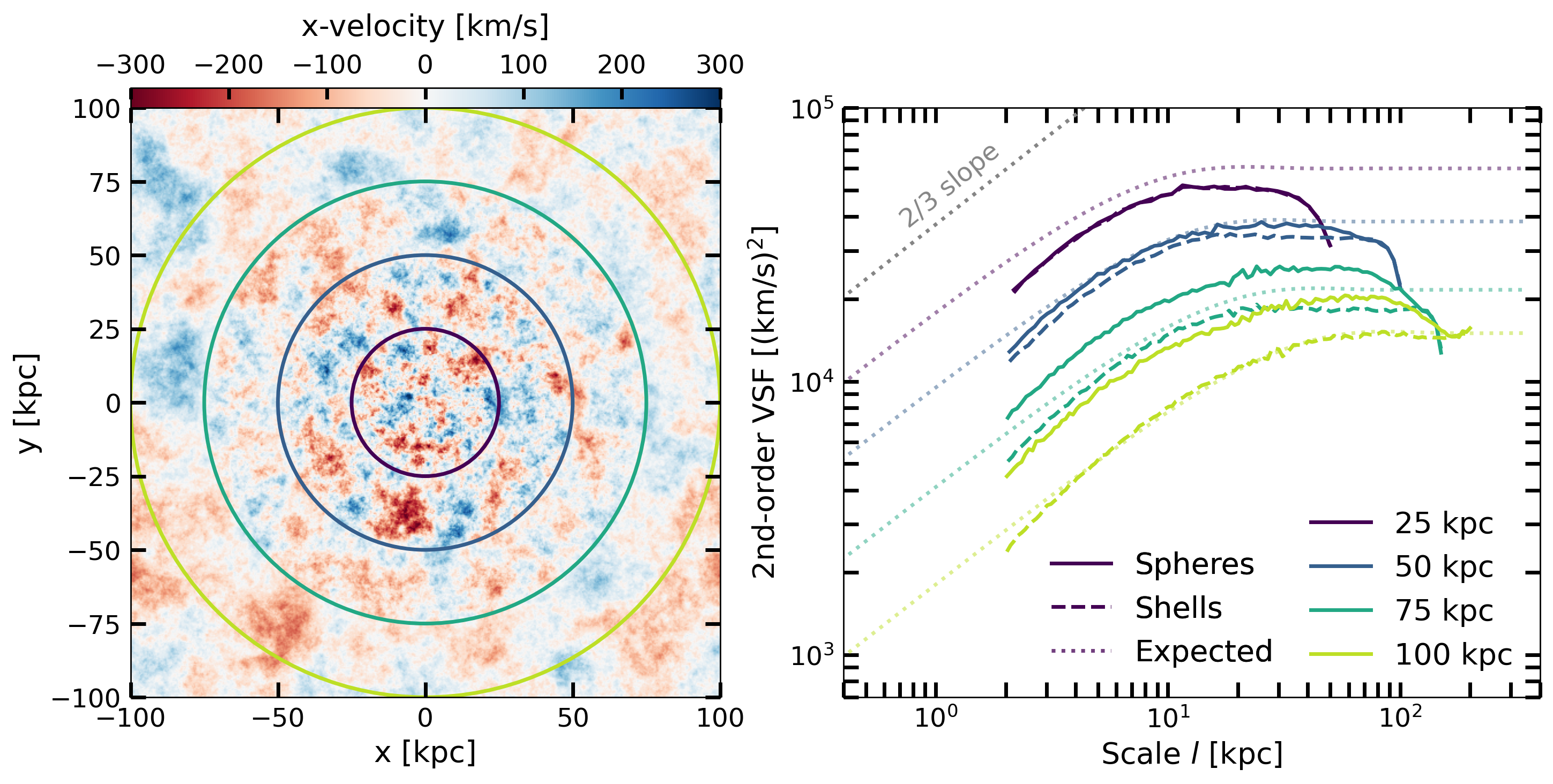}
    \caption{\emph{(Left):} A slice of the $x$-velocity field through the center of a box generated with idealized Kolmogorov turbulence, where both $k_\mathrm{peak}$ and $\sigma_v$ decrease in radial shells from the inside out. Colored circles mark the edges of the radial bins where transitions in $k_\mathrm{peak}$ and $\sigma_v$ occur. \emph{(Right):} The expected (dotted) and measured VSFs from the radial shells (dashed) or from filled spheres (solid) of the same radii as the shells. The shell VSFs match expectations closely, as they have minimal contamination from different $k_\mathrm{peak}$ and $\sigma_v$ outside the shell, but the sphere VSFs can be quite different due to mixing of $k_\mathrm{peak}$ and $v_\sigma$ values in the measurement region.}
    \label{fig:ideal_diffturb}
\end{figure*}

In practice, we generate 4 boxes, with $k_\mathrm{peak}=(8,6,4,2)$ and $\sigma_v=(100,80,60,50)$ km/s. We then combine these boxes by taking the velocity fields of each only within certain radii from the center. Between $r=0$ kpc and $r=25$ kpc, the velocity fields are taken from the box with $k_\mathrm{peak}=8$ and $\sigma_v=100$ km/s; between $r=25$ kpc and $r=50$ kpc, the velocity fields are taken from the box with $k_\mathrm{peak}=6$ and $\sigma_v=80$ km/s; and so on for $r=50$--75 kpc and $r>75$. We then smooth the transitions between shells by weighting each shell's contribution to every cell by a Gaussian that is centered on the center of the radial bin with a standard deviation of $1/4$ of the width of the radial bin. The left panel of Figure~\ref{fig:ideal_diffturb} shows a slice of the $x$-velocity field resulting from this process, with circles denoting the transitions between each radial shell. The $y$- and $z$-velocity fields are generated the same way. We choose to generate the turbulent velocity fields in this way so that this example closely matches the VSFs measured in radial shells in the FOGGIE simulations (Figure~\ref{fig:radial_bins}).

We then measure the VSF from the radial shells and from filled-in spheres of the same radius. The shell VSFs contain minimal contamination from other $k_\mathrm{peak}$ and $v_\sigma$ values with smaller radii than the shell, but the sphere VSFs include information from all cells within their radius, and are thus contaminated by the smaller-radius shells. The right panel of Figure~\ref{fig:ideal_diffturb} shows the VSFs expected from the $E(k)$ spectrum that generated the velocity field in each shell (dotted) as compared to the VSFs measured from the shells (dashed) or from the spheres of the same radii (solid).

The shell VSFs follow the expectation closely, with some deviation due to the Gaussian smoothing at the edges of the shells that mixes in a small amount of turbulence of different $k_\mathrm{peak}$ and $\sigma_v$ values. This shows that as long as the turbulence is consistent across the shell, restricting the information used to measure the VSF to a radial shell does not impact the accuracy of the measured VSF.

The sphere VSFs, on the other hand, begin to deviate strongly from both the expectation and the shell VSFs in the outer two radii (75 and 100 kpc), where the VSFs are measured from turbulence of a variety of $k_\mathrm{peak}$ and $\sigma_v$ values. The turnover or flattening locations measured by the sphere VSFs are surprisingly similar to their corresponding shells, suggesting that the flattening scale is dominated by the largest driving scale, or smallest value of $k_\mathrm{peak}$ \citep[consistent with the findings of][]{Yoo2014}. The normalization of a 3D VSF is set by $6\sigma_v^2$, and the larger spheres contain a variety of $\sigma_v$ values. Because of this, the normalization of a given sphere is equal to a volume-weighted average of $\sigma_v$ within the sphere's radius. For example, the $r=0$--25 kpc shell contributes $\sim1\%$ of the total volume to the outermost (100 kpc) sphere, the $r=25$--50 kpc shell contributes $\sim11\%$, the $r=50$--75 kpc contributes $\sim30\%$, and the $r=75$--100 kpc shell contributes the remaining $\sim58\%$. The 100 kpc sphere has a normalization at the flattening scale of $2\times10^4$ $(\mathrm{km}/\mathrm{s})^2$, which is equivalent to $6\times(0.01\times100^2 + 0.11\times80^2 +0.30\times60^2+0.58\times50^2)$.

In summary, VSFs still contain valuable information even when they are measured from turbulence with a variety of driving scales and velocity dispersions. The flattening scale of the VSF appears to trace the largest driving scale, while the normalization of the VSF traces a volume-weighted average velocity dispersion. This is of course a very idealized case, but the similarity of the right panel of Figure~\ref{fig:ideal_diffturb} to Figure~\ref{fig:radial_bins} suggests that the VSFs measured in radial bins within FOGGIE are in fact capturing a shifting of the driving scale to larger scales and a shifting of the turbulent velocity dispersion to smaller values as the distance from the central galaxy increases.

\section{Geometric Effects of Projection on the VSF}
\label{appdx:projection_ladder}

\begin{figure*}
    \centering
    \includegraphics[width=0.49\linewidth]{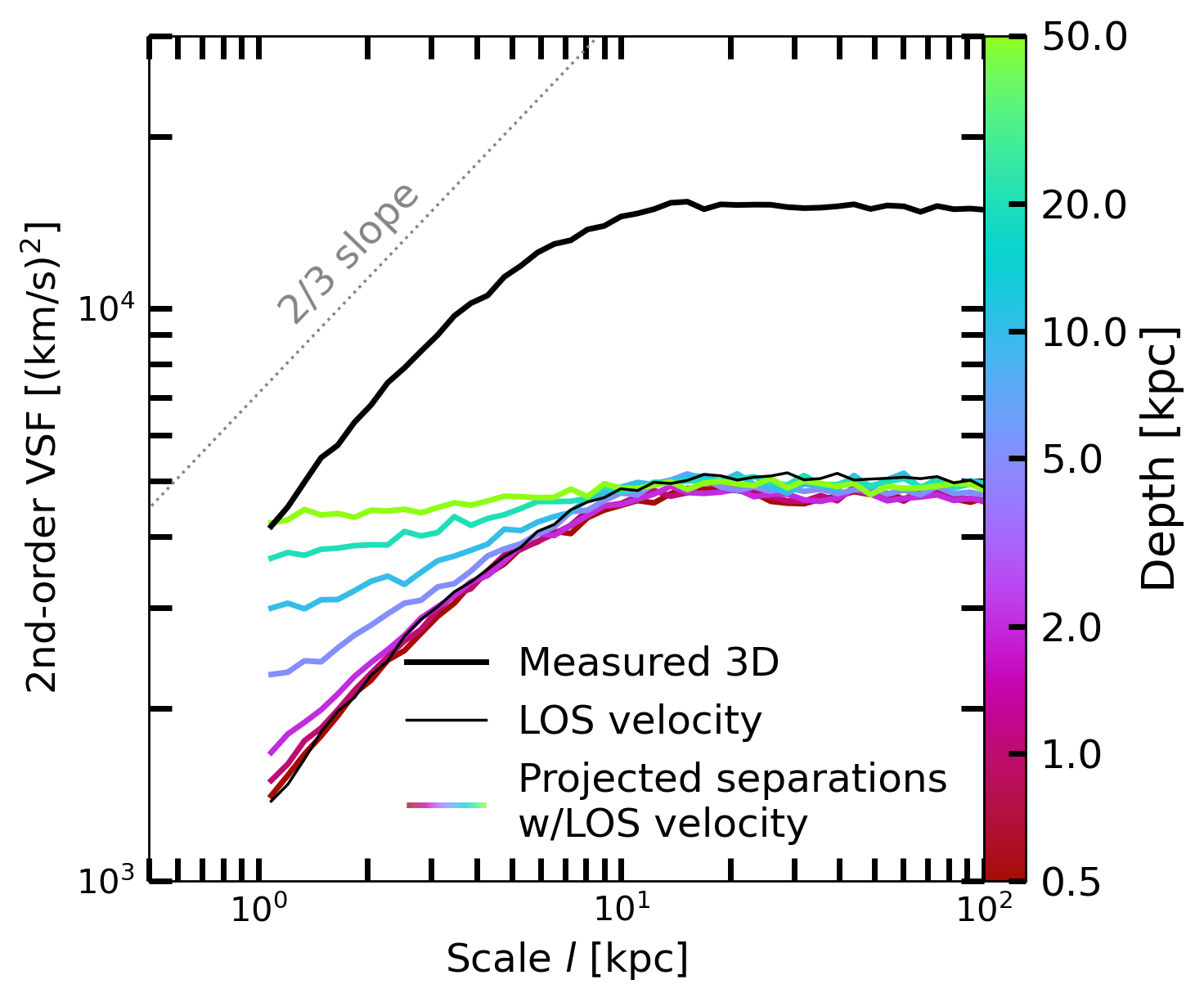}
    \includegraphics[width=0.49\linewidth]{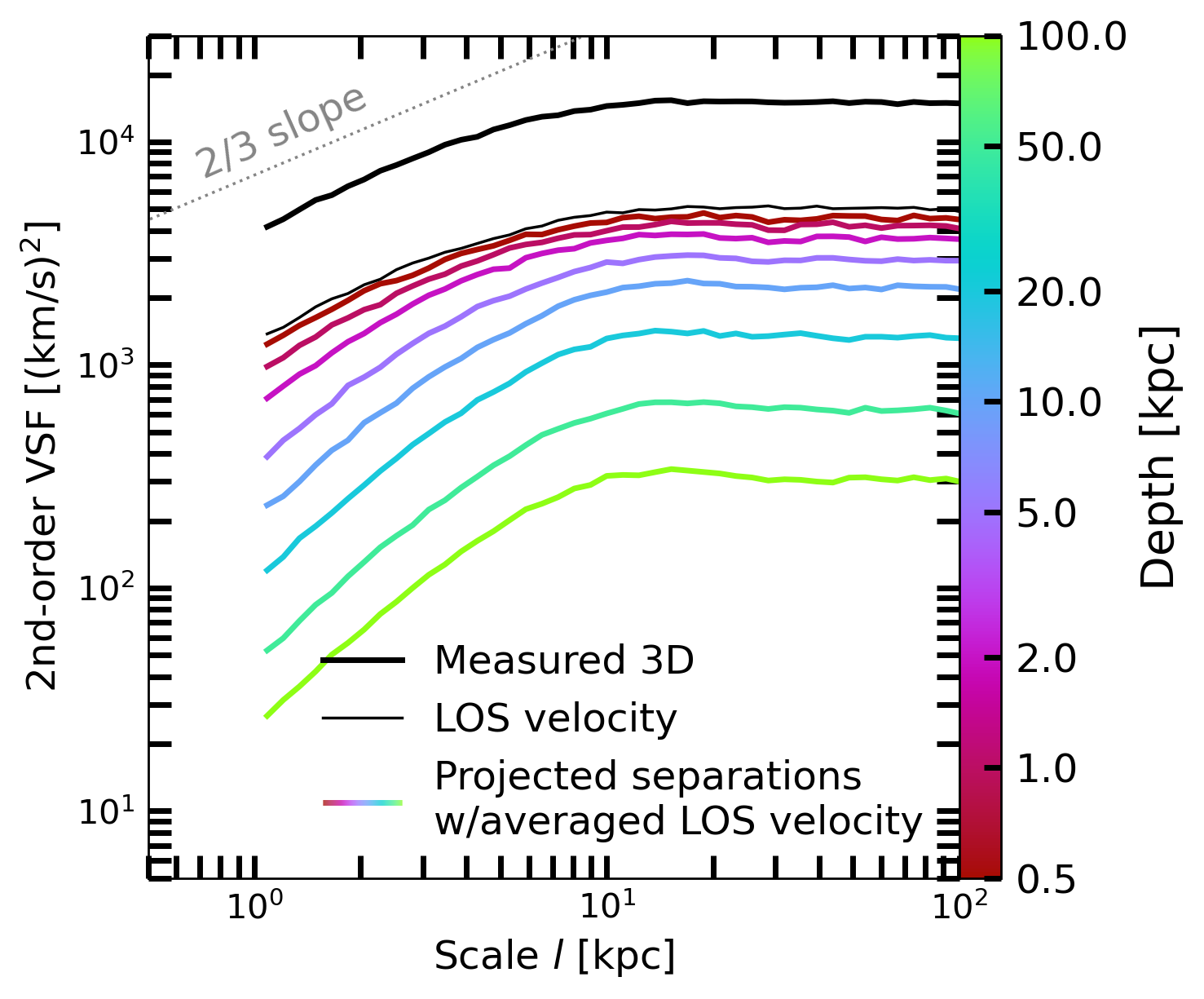}
    \caption{The second-order VSFs resulting from different rungs of the projection ladder (see text). \emph{(Left):} The first two rungs of the ladder. The thick black curve shows the full 3D VSF, while the thin black curve shows the VSF calculated with 3D positions but just one dimension of the velocity, along the line of sight. The colored curves show VSFs calculated using 2D projected positions for each point in the pair, with the LOS velocity for each point chosen randomly from somewhere along the LOS up to a certain depth, as indicated by the color coding. \emph{(Right):} The third rung of the ladder. The thick and thin black curves are the full 3D VSF and a VSF with 3D positions but only LOS velocity, as in the left panel. The colored curves show VSFs calculated with 2D projected positions for each point in the pair and velocities calculated as the average of all LOS velocities up to some depth, as indicated by the color coding.}
    \label{fig:appdx_projection_ladder}
\end{figure*}

We use the idealized turbulent velocity fields generated in Appendix~\ref{appdx:box_size} to investigate each step of the projection ``ladder" used in Section~\ref{subsec:proj_effects}. Because the idealized velocity fields are perfectly isotropic with a known turbulent spectrum and do not contain any non-turbulent gas flows, they can be used to validate the impact of projection effects on the VSF as found in Section~\ref{subsec:proj_effects} for the cosmological simulations. Figure~\ref{fig:appdx_projection_ladder} shows the effect of the rungs of the ladder as we project the 3D VSF in the idealized velocity fields to a 2D mock-observed VSF.

The first ladder rung is simply reducing the dimensionality of the velocity field to only the LOS velocity. In the perfectly isotropic velocity fields, this trivially reduces the VSF to one-third its fully 3D value, as shown by the thin black curves in both panels of Figure~\ref{fig:appdx_projection_ladder}. The thick black curves in both panels show the fully 3D VSF measured from the idealized velocity fields, for reference.

The second rung is to reduce the dimensionality of the calculated separations between pairs of points as well, by using 2D projected separations instead of the 3D separation, and continue using the LOS velocity. This effectively removes the information of the depth of a point in a pair along the LOS. We test the case wherein the depth of a point is not known precisely, but is known to be restricted to a specific range in depth, in the colored curves of the left panel of Figure~\ref{fig:appdx_projection_ladder}. To calculate this, we first reduce the cubic velocity fields to a slab of the specified depth, randomly select a large number of pairs of points as before, and calculate the VSF using 2D projected separations between each pair and only the LOS velocity of each point in the pair. We repeat this process for depths of the slab of 0.5, 1, 2, 5, 10, 20, and 50 kpc, and color each VSF curve by the depth of the slab from which it was calculated. The driving scale of the turbulence in this case is set to 25 kpc, which places the turnover in the VSF at 12.5 kpc. For very small depths of 0.5, 1, or 2 kpc (i.e., $\lesssim 0.2\times$ the flattening scale), the VSF is not much different from the thin black curve (representing 3D separations and LOS velocity). However, as the depth of the slab increases, representing more confusion in the location of each point along the LOS, the VSF below the flattening scale becomes significantly more shallow. At very large depths greater than the flattening scale, the VSF approaches a completely flat line, washing out all information other than the value of the velocity dispersion in the box.

The third ladder rung is to select random pairs of points only in the 2D projected space, and use the average of the LOS velocity for all cells along each point's full line of sight. Because the idealized turbulent velocity fields do not have an associated density with which to calculate a density-weighted average, we calculate only volume-weighted averages and this is the final rung. We again perform the experiment using slabs of different depths, and show the result in the colored curves of the right panel of Figure~\ref{fig:appdx_projection_ladder}. We use the same values of the depth as for ladder rung two, but include an additional 100 kpc depth, which is the depth of the entire generated velocity field box. The thick and thin black curves in the right panel are equivalent to those in the left panel, for reference. Again, we see that very small values of the depth do not much affect the VSF, but calculating the VSF for large depths requires averaging the LOS velocity over a large number of values, which reduces the overall normalization because the average LOS velocity approaches zero (by construction, because the average velocities in the generated velocity fields are set to zero). In addition to decreasing the normalization, the slope of the VSF below the flattening point also becomes steeper as the depth of the velocity averaging increases, in agreement with similar results found by \citet{Xu2020}.

In general, we find the same trends in the multi-step projection ladder with the idealized turbulent velocity fields as we found for the cosmological simulation in Sec.~\ref{subsec:proj_effects}. The depth of the selected region of the cosmological simulation is quite large (100 kpc), but the density-squared weighting reduces the effective depth from which the 2D VSF is calculated to only the region of higher gas density near the galaxy. Because of this, the information in the 2D VSF is not completely wiped out, and the measured density-squared-weighted projected VSF has a slope and a normalization between those of ladder rungs one and three (see Figure~\ref{fig:projection_ladder}).

\bibliographystyle{aasjournal}
\bibliography{bibliography.bib}

\end{document}